\documentclass[12pt]{article}
\usepackage{amsmath, latexsym, amssymb, amscd, float,slashed}
\usepackage{tikz}
\usetikzlibrary{arrows,shapes}
\usetikzlibrary{trees}
\usetikzlibrary{matrix,arrows} 				
\usetikzlibrary{positioning}				
\usetikzlibrary{calc,through}				
\usetikzlibrary{decorations.pathreplacing}  
\usepackage{pgffor}							

\usetikzlibrary{decorations.pathmorphing}	
\usetikzlibrary{decorations.markings}
\usetikzlibrary{shapes,arrows,positioning,automata,backgrounds,calc,er,patterns}

\restylefloat{table}
\newcommand{\C}{\mathbb{C}}
\newcommand{\R}{\mathbb{R}}
\newcommand{\N}{\mathbb{N}}
\newcommand{\newew}{U_{EW}(2)}
\newcommand{\oldew}{SU_I(2)\times U_Y(1)}

\newcommand{\D}{\not\!\!{D}}

\newcommand{\bs}[1]{\mathbf{#1}}
\newcommand{\Bold}[1]{\mbox{\boldmath$\mathit{#1}$}}
\newcommand{\sBold}[1]{\mbox{\boldmath$\mathit{\scriptstyle{#1}}$}}
\newcommand{\ssBold}[1]{\mbox{\boldmath$\mathit{\scriptscriptstyle{#1}}$}}
\newtheorem{remark}{Remark}[section]
\tikzset{
    vector/.style={decorate, decoration={snake}, draw},
	provector/.style={decorate, decoration={snake,amplitude=2pt}, draw},
	antivector/.style={decorate, decoration={snake,amplitude=-2pt}, draw},
    fermion/.style={draw=black, postaction={decorate},
        decoration={markings,mark=at position .55 with {\arrow[draw=black]{>}}}},
    fermionbar/.style={draw=black, postaction={decorate},
        decoration={markings,mark=at position .55 with {\arrow[draw=black]{<}}}},
    fermionnoarrow/.style={draw=black},
    gluon/.style={decorate, draw=black,
        decoration={coil,amplitude=4pt, segment length=5pt}},
    scalar/.style={dashed,draw=black, postaction={decorate},
        decoration={markings,mark=at position .55 with {\arrow[draw=black]{>}}}},
    scalarbar/.style={dashed,draw=black, postaction={decorate},
        decoration={markings,mark=at position .55 with {\arrow[draw=black]{<}}}},
    scalarnoarrow/.style={dashed,draw=black},
    electron/.style={draw=black, postaction={decorate},
        decoration={markings,mark=at position .55 with {\arrow[draw=black]{>}}}},
	bigvector/.style={decorate, decoration={snake,amplitude=4pt}, draw},
}

\usepackage{amsmath, latexsym, amssymb, amscd}

\begin{document}

\title{A non-Standard Standard Model}
\author{J. LaChapelle}

\maketitle

\begin{abstract}
This paper examines the Standard Model under the strong-electroweak symmetry group $SU_S(3)\times U_{EW}(2)$ subject to the Lie algebra condition $\mathfrak{u}_{EW}(2)\not\cong \mathfrak{su}_{I}(2)\oplus \mathfrak{u}_{Y}(1)$. Physically, the condition ensures that all electroweak gauge bosons interact with each other prior to symmetry breaking --- as one might expect from $U(2)$ invariance. This represents a crucial shift in the identification of physical gauge bosons: Unlike the Standard Model which posits a change of Lie algebra basis induced by spontaneous symmetry breaking, here the basis is unaltered and $A,\,Z^0,\,W^\pm$ represent the physical bosons both before and after spontaneous symmetry breaking.

Our choice of $\mathfrak{u}_{EW}(2)$ requires some modification of the matter field representation of the Standard Model. The group $U_{EW}(2)$ admits two pertinent defining representations, ${\mathbf{2}}$ and its $U(2)$-conjugate ${\mathbf{2^c}}$, related by a large gauge transformation. Accordingly, the product group structure calls for strong-electroweak degrees of freedom in the $(\mathbf{3},\mathbf{2})$ and the $(\mathbf{3},{\mathbf{2^c}})$ of $SU_S(3)\times U_{EW}(2)$ that possess integer electric charge just like leptons. These degrees of freedom play the role of quarks, and they lead to a modified Lagrangian that nevertheless reproduces transition rates and cross sections equivalent to the Standard Model. In particular, they reproduce the fractional electric charge of quark currents.

The close resemblance between quark and lepton electroweak doublets in this picture suggests a mechanism for a speculative phase transition between quarks and leptons that stems from the product structure of the symmetry group. Our hypothesis is that the strong and electroweak bosons see each other as a source of decoherence. In effect, lepton representations get identified with the $SU(3)$-trace-reduced quark representations. This mechanism allows for possible extensions of the Standard Model that don't require large inclusive multiplets of matter fields.
\end{abstract}

\section{Introduction}
The present-day Standard Model (SM) started out as an electroweak (EW) theory of leptons.\cite{G1,W1,S} Only later were hadrons tentatively incorporated by considering
the known structure of charged hadronic currents. This led to a
postulated hadronic composition by quarks along with their assumed weak isospin
and hypercharge quantum numbers.\cite{GIM,FGL,W2}

In the quark sector of the fledgling model, the canonical status enjoyed by isospin $I$ and hypercharge $Y$ was a consequence of the of
the EW symmetry group $\oldew$ and the success of the
Gell-Mann/Nishijima relation ($Q\propto I+1/2Y$) in classifying
mesons and baryons in various approximate $SU(2)$ and $SU(3)$ flavor
symmetry models. Historically, this led to the conclusion that the
$(u,d,s)$ quarks possess fractional electric charge. But it was soon recognized that \textcolor{green}{fermion} statistics required additional degrees of freedom, and the idea of a local three-fold color symmetry group was born. Consequently, adding gauged $SU_C(3)$, still assuming fractional electric
charge, and using the Gell-Mann/Nishijima relation led
to the assignment of quarks in the
$(\mathbf{3},\mathbf{2},\mathbf{1/3})$ representation of gauged
$SU_C(3)\times\oldew$.

Our aim is to study how this historical picture changes if  we start instead with the symmetry group $SU_S(3)\times U_{EW}(2)$.\footnote{The subscript $S$ stands for `strong' and $EW$ stands for `electroweak'. We will explain the significance of the distinction between the subscripts $S,\,EW$ v.s. $C,\,I,\,Y$ later.} There are good reasons to believe the
EW group is $U(2)$. First, if $\rho$ is the representation
of $SU(2)\times U(1)$ furnished by the lepton fields of the SM,
then $\mathrm{ker}\rho=Z_2$ and therefore the lepton matter fields
do not furnish a faithful representation.\cite{I} The group
that does act effectively on all  matter fields is $(SU(2)\times
U(1)) /Z_2=U(2)$. Second, both $SU(2)\times U(1)$ and $U(2)$ have
the same covering group $SU(2)\times\R$. Representations of $SU(2)\times\R$
will descend to representations of $SU(2)\times U(1)$ or $U(2)$
if the associated discrete factor groups are represented trivially by the unit matrix. For $SU(2)\times U(1)$ this
requirement implies no relationship between isospin and
hypercharge, but for $U(2)$ it implies $n=I+1/2Y$ with $n$ integer (\cite{GIL}, pg. 145). Identifying $n$ with electric charge renders the
Gell-Mann/Nishijima relation and electric charge quantization a
\emph{consequence} of $U(2)$. Third, symmetry reduction from
$U(2)$ to $U(1)$ is less constrained and more natural than from
$SU(2)\times U(1)$.\cite{I,D} Fourth, to the extent that a fiber bundle formulation over a paracompact base space is an appropriate physical model, the most general structure group for a complex matter field doublet in this case is $U(2)$.\cite{LA2} The point is that $\mathfrak{su}(2)\oplus \mathfrak{u}(1)\cong \mathfrak{u}(2)$ as vector spaces, but as groups $SU(2)\times U(1)\not\cong U(2)$. This subtle difference in symmetry groups has a big effect on the phenomenologically apposite basis of the associated Lie algebra that represents gauge fields.

The first task is to determine this basis. On physical grounds we demand that: 1) the gauge bosons are associated with a particular (up to group conjugation) `physical basis' in a real subalgebra of $\mathfrak{u}(2,\C)$ that is endowed with a suitable $Ad$-invariant inner product (which in our case differs from the Killing inner product); and 2) \emph{all} gauge bosons are allowed to take part in boson--boson interactions --- as intuition suggests befits a $U(2)$ symmetry group. These conditions restrict the starting gauge symmetry to a subgroup that we denote by $U_{EW}(2)\subset U(2,\C)$. A key feature of its Lie algebra $\mathfrak{u}_{EW}(2)$ is that its generators do not form a basis for the algebra decomposition $\mathfrak{su}_{I}(2)\oplus \mathfrak{u}_{Y}(1)$ of the SM. Instead, they comprise the `physical basis' associated with the $A,\,Z^0,\,W^\pm$ gauge bosons. This basis is supposed physically proper both before and after spontaneous symmetry breaking (SSB). Implementing this standpoint calls for some fairly moderate modifications of the SM.

To understand the motivation and implications of these modifications, in $\S\,2$ we embark on a somewhat pedestrian review of the adjoint and defining representations of Lie algebras of product groups. We propose physical gauge bosons characterized by conserved charges should be identified  with a distinguished Cartan basis with its concomitant root system. This Cartan basis of gauge bosons is `physical' in the sense that one can identify its roots/quantum numbers with the conserved charges associated with coupling constants parametrizing the Lie algebra inner product. We confirm that quantization and renormalization don't spoil this identification, and so it is permitted to use the terms `root', `quantum number', and `charge' interchangeably. (To be precise, the term `charge' here refers to the eigenvalue of a charge operator acting on a gauge field.) Demanding all generators of the Cartan subalgebra be gauge equivalent determines the generators of $\mathfrak{u}_{EW}(2)$ and their associated charges.

In this Cartan-basis picture, in which the neutral gauge bosons span the Cartan subalgebra, elementary matter fields are identified with eigenfields of the neutral gauge bosons in the defining representation whose charges are determined by their associated weight system. This guarantees that matter fields and gauge bosons exchange the same types of charges --- which are ultimately determined by the Cartan basis. Due to the product nature of the symmetry group, the eigenfields associated with neutral $SU_S(3)\times U_{EW}(2)$ gauge bosons evidently include \emph{both} representations $(\mathbf{3},\mathbf{2})$ and $(\mathbf{3},{\mathbf{2^c}})$ where ${\mathbf{2^c}}$ is $U_{EW}(2)$-conjugate via a ``large'' gauge transformation; one that is not homotopic to the identity. This means that $(\mathbf{3},\mathbf{2})$ and $(\mathbf{3},{\mathbf{2^c}})$ each carry two types of conserved strong charge and two types of EW charge; the latter of which only one is conserved after SSB and corresponds to an \emph{integer} unit of electric charge characterized by the (neutral) photon --- in glaring contrast to quarks.

But, since $(\mathbf{3},\mathbf{2})$ and $(\mathbf{3},{\mathbf{2^c}})$ are supposed to be the building blocks of hadrons, an apparent contradiction arises: How can a triplet of these hadronic constituents (HC), each of which possess integer electric charge, combine to
form hadrons with their observed electric charges? We show in the remainder of $\S\,2$
that the charge carried by a matter
eigenfield versus the coupling strength coming from its associated current   are not necessarily equivalent.

How can this be? Since the defining representation $\mathbf{2}$ of $U_{EW}(2)$ and its $U_{EW}(2)$-conjugate ${\mathbf{2^c}}$  are related by a large gauge transformation, they exhibit homotopically inequivalent representations. The most general (quadratic with minimal coupling) effective Lagrangian that reflects this fact contains a linear combination of HC kinetic terms representing both $(\mathbf{3},\mathbf{2})$ and $(\mathbf{3},{\mathbf{2^c}})$. We will see that the ratio of scalar factors for the two terms is non-trivial: it can't be absorbed in the normalization, it doesn't get renormalized, and it doesn't spoil the gauge symmetry. A non-trivial ratio renders scaled coupling constants that ultimately become coupling strengths in matter field currents. Consequently the charge is not always equivalent to the coupling strength between currents and gauge bosons.

This is an inceptive observation, and it inspires a relatively mild variation of the SM. With the physically relevant gauge bosons and their matter eigenfields in hand, we begin the explicit construction  of this non-Standard Model (n-SM) in $\S\,3$. After determining the defining representations of $U_{EW}(2)$ and $SU_S(3)$, a Lagrangian density is proposed. We stress that the only pieces that differ from the SM are the quark contributions --- our's include a linear combination of both $(\mathbf{3},\mathbf{2})$ and $(\mathbf{3},{\mathbf{2^c}})$ whose scalar factors are \emph{a priori} free parameters, but anomaly cancellation fixes them.

As a result, the strong and electroweak currents coming from the n-SM agree with those from the SM \emph{provided quark currents are identified with a pair of HC currents}. In particular, the electromagnetic (EM) current contains the expected
$2/3e$ and $-1/3e$  coupling strengths even though the HC have integer
 electric charge.  For this reason, we call the HC ``iquarks". Essentially, a SM quark doublet is interpreted as an average description of two n-SM iquark doublets. Insofar as experiments are not able to distinguish individual iquark currents, we prove \emph{spin averaged transition rates and cross sections of the n-SM coincide with those of the SM}.

Quarks with integer electric charge have been
proposed before (see, e.g., \cite{GIM, PS}, and the review
of \cite{GS}). However, the symmetry groups of these models are
not the standard model symmetry group $SU_C(3)\times SU_I(2)\times U_Y(1)$,  and the n-SM presented here is not related to these models. Also, the
proposed iquarks are not ``preons'' or ``pre-quarks'' (see, e.g.,
\cite{T1,PS2,AH} and the review of \cite{T2}). That is,
conventional quarks are not composite states of the iquarks.
Instead, within this framework, conventional quarks can be
interpreted as a certain \emph{superposition} of the iquarks.

An important consequence of including both $(\mathbf{3},\mathbf{2})$ and $(\mathbf{3},{\mathbf{2^c}})$ in the Lagrangian density is the presence of an approximate global $(U(2)_L\times U(2)_R)\times\mathbb{Z}_2$ symmetry in the limit of vanishing iquark masses. If this global symmetry is broken down to $SU(2)_V\times U(1)_V$ by an $SU_S(3)$ condensate, the resulting pseudo-Goldstone boson is a charged $SU(2)$ doublet. As such, it is a contender for a Higgs boson that could be immune to the hierarchy problem.\cite{G2,CO} We will not address this issue here but leave it to future work.

Perhaps the primary benefit of the n-SM is that it affords an alternative perspective for extensions. In the final section $\S\,4$, we hypothesize the possibility of a phase transition induced by the product group structure. Our idea is that gauge bosons of $SU(3)$ versus $U(2)$ see themselves as a quantum system and the other as an environment through their mutual interactions with matter fields. As a consequence, in restricted regions of phase space, their mutual decoherence is posited to precipitate a transmutation of matter field representations. In particular, the affected defining representation gets reduced to its trace due to decoherence. We interpret this representation transmutation as a phase transition, and it occurs without breaking symmetry. It is a non-perturbative dynamical effect not modeled by the Lagrangian; so our n-SM is an effective theory at the EW energy scale valid only in certain regions of phase space. If this mechanism is viable, it allows otherwise excluded options for unifying symmetry groups, because iquarks and leptons are not required to belong to inclusive multiplets.

It should be mentioned that similarities between quarks and leptons have long been noticed. Besides inspiring ideas of unifying groups, they have prompted some attempts to endow leptons with an $SU(3)$ symmetry (see e.g. \cite{FV}),  and they are getting attention in recent years under the program of quark-lepton complementarity which studies phenomenological similarities between the generation-mixing matrices \cite{XZ}.

\section{Intrinsic v.s. Extrinsic Charge}\label{section 1}

\subsection{Kinematical quantum numbers}
Begin with a gauge field theory endowed with an internal symmetry group
that is a direct product group $G=G_1\times\cdots\times
G_n=:\times_n\,G_i$ where $n\in\N$ and the $G_i$ with
$i\in\{1,\ldots,n\}$ are Lie groups that mutually commute.
Associated with each subgroup $G_i$ is a Lie algebra
$\mathcal{G}_i$ with basis
$\{\bs{g}_{a_{i}}\}_{a_i=1}^{\mathrm{dim}G_i}$. The full Lie
algebra is $\mathcal{G}:=\oplus_n \mathcal{G}_i$ (in obvious
notation). Recall that the Lie algebra does not uniquely determine
the Lie group.

Given that a physical system is invariant under the associated
gauge group, it is possible to deduce some general properties
or attributes of the associated gauge and matter fields based
solely on the mathematics of the symmetry group and its
representations \cite{COR,{GIL}}. In particular, the mathematics identifies distinguished
bases and associated eigenvalues (which we will identify with quantum numbers) in the vector
spaces furnishing the adjoint representations of each $\mathcal{G}_i$.

The distinguished bases are physically important in the sense that they describe the adjoint-representation counterpart of matter field irreducible representations. As such, they model the concepts of particle/antiparticle and particle--particle interaction with charge exchange.
For local symmetries, these
special bases can be chosen independently at each spacetime point, essentially
creating an unchanging structure by which to associate the
unchanging quantum numbers of elementary particles --- both bosons and fermions.

It is well-known, of course, that the distinguished bases are induced by the Cartan subalgebra. This is a well-worn story. But, as we are dealing with a product group $G=\times_n\,G_i$, we will spend a moment repeating it as a means of review and to set some notation.

Consider the adjoint representation $ad:\mathcal{G}_i\rightarrow
GL(\mathcal{G}_i)$ on
 the \emph{complex extension} of $\mathcal{G}_i$. For a given element $c^{a_i}\bs{g}_{a_i}$ (with
$c^{a_i}\in \C$) in the Lie algebra, the adjoint representation
yields a secular equation
\begin{equation}\label{secular}
  \prod_{k_i=0}^{r_i}(\lambda-\alpha_{k_i})^{d_{k_i}}=0
\end{equation}
where $\alpha_{k_i}$ are roots of the secular
equation with multiplicity $d_{k_i}$ and $r_i$ is the rank of $\mathcal{G}_i$. Since $\lambda=0$ is always
a solution, we put $\alpha_0=0$. Note that
$\sum_{k_i=0}^{r_i}d_{k_i}=\mathrm{dim}G_i$. Associated with the
roots $\alpha_{k_i}$ (which may not all be distinct in general)
are $r_i$ independent eigenvectors.

The roots and their associated eigenvectors determine the Jordan block form of the element
$ad(c^{a_i}\bs{g}_{a_i})$. That is, there exists a non-singular
transformation of $ad(c^{a_i}\bs{g}_{a_i})$ into Jordan canonical
form. With respect to the Jordan canonical form, the vector space
that carries the representation $ad(\mathcal{G}_i)$ decomposes into a direct sum of subspaces:
\begin{equation}\label{decomposition}
\mathcal{G}_i=\oplus_{k_i} V_{\alpha_{k_i}}
\end{equation}
with each $V_{\alpha_{k_i}}$ containing one eigenvector and
$\mathrm{dim}V_{\alpha_{k_i}}=d_{k_i}$. This decomposition is with respect to any given element in the Lie algebra.

Now, \emph{regular} Lie algebra elements are defined by the conditions that:
(i) they lead to a decomposition that maximizes the distinct
roots $\alpha_{k_i}$ (equivalently, minimize the dimension of
$V_{\alpha_{k_i}}$), and (ii) they all determine the same
$V_{0_i}$. For decomposition associated with regular elements, the
subspaces $V_{\alpha_{k_i}}$ have potentially useful properties
for describing physical gauge bosons:
\begin{itemize}
\item{$[V_{0_i},V_{0_i}]\subseteq V_{0_i}$ and hence $V_{0_i}$ is a subalgebra.
It is known as a Cartan subalgebra.}
\item{The subspace $V_{0_i}$ carries a representation of the Cartan subalgebra.
Since its rank is $0$, the Cartan subalgebra is solvable; in fact
nilpotent.}
\item{$[V_{0_i},V_{\alpha_{k_i}}]\subseteq V_{\alpha_{k_i}}$. Hence, each
$V_{\alpha_{k_i}}$ is invariant with respect to the action of
$V_{0_i}$ and so carries a representation for $V_{0_i}$. Moreover,
since $V_{0_i}$ is solvable, it has a simultaneous eigenvector
contained in $V_{\alpha_{k_i}}$. More specifically, associated
with the secular equation for an element of the subalgebra
$V_{0_i}$ with basis $\{\bs{h}_{s_i}\}_{s_i=1}^{d_{0_i}}$ is a set
of $\mathrm{dim}V_{0_i}=d_{0_i}$ roots, collectively denoted by
$\Bold{q}_i:=({q_1}_i,\ldots,{q_{d_0}}_i)$, and a corresponding
eigenvector $\bs{e}_{{\alpha}_{k_i}}\in V_{\alpha_{k_i}}$ such
that
\begin{equation}\label{eigenvalue}
  [\bs{h}_{s_i},\bs{e}_{{\alpha}_{k_i}}]=q_{s_i}\bs{e}_{{\alpha}_{k_i}}\;,
\end{equation}
or more succinctly,
\begin{equation}\label{eigenvalues}
 [\bs{h}_i,\bs{e}_{{\alpha}_{k_i}}]=\Bold{q}_{i}\bs{e}_{{\alpha}_{k_i}}\;,
\end{equation}
 In particular, this holds for $\alpha_0=0$. That is, there exists
 an $\bs{e}_{{0}_i}\in V_{0_i}$ such that
\begin{equation}\label{zero}
[\bs{h}_i,\bs{e}_{{0}_i}]=\bs{0}\;.
\end{equation}
 }
\item{If $V_{0_i}$ is contained in the derived algebra of
$\mathcal{G}_i$, then for $V_{\alpha_{k_i}}$, there is at least
one $V_{\beta_{k_i}}$ such that
$[V_{\alpha_{k_i}},V_{\beta_{k_i}}]\subseteq V_{0_{k_i}}$. This
implies that, for  $\Bold{q}_i$ associated with each
$\bs{e}_{\alpha_{k_i}}$, there is at least one
$\bs{e}_{\beta_{k_i}}$ with roots $-\Bold{q}_i$. Additionally, any
$\Bold{q}'_i\neq -\Bold{q}_i$ must be a rational multiple of
$\Bold{q}_i\neq 0$. }
\end{itemize}

These properties can be used to characterize  `physical' gauge
bosons if we make one restriction: for $\alpha_{k_i}\neq 0$,
$\mathrm{dim}\oplus_{\alpha_{k_i}} V_{\alpha_{k_i}}=\mathrm{dim}G_i-d_{0_i}=r_i$.
That is $\mathrm{dim}V_{\alpha_{k_i}}=1$ for all $\alpha_{k_i}\neq
0$. Without this restriction, there would be no means
(mathematically) to distinguish between basis elements, and hence
gauge bosons, in a given $V_{\alpha_{k_i}}$. As a consequence of
this restriction, we must have $[V_{0_i},V_{0_i}]=0$ since
otherwise $[[V_{0_i},V_{0_i}],V_{\alpha_{k_i}}]$ in the Jacobi
identity leads to a contradiction.

The commutativity of $V_{0_i}$ is a necessary condition for
$\mathcal{G}_i$ to be the direct sum of
one-dimensional abelian and/or simple algebras. Moreover,
eventually the Lie algebra elements will be promoted to quantum
fields so the adjoint carrier space is required to be Hilbert.
Therefore, the inner product on the complex Lie algebra (or subspace thereof) is required to be
Hermitian positive-definite. This implies that candidate symmetry group Lie algebras \emph{may} be the direct sum
of $\mathfrak{u}(1)$ and/or \emph{compact} simple algebras since then the Killing inner product is positive-definite. However, it is consistent to allow a slightly larger class of algebras --- reductive to be specific --- provided they are endowed with $Ad$-invariant Hermitian inner products and commuting Cartan subalgebras.

It should be kept in mind that the class of Lie algebras under
consideration up to now have been \emph{complex}. However, the adjoint representation is real so the gauge bosons' kinematical quantum numbers are real. Since we want the matter fields to be characterized by the same real quantum numbers and we want to distinguish between field/anti-field, then on physical grounds \emph{we insist on symmetry groups generated by real, reductive Lie algebras endowed with a suitable $Ad$-invariant inner product} --- or subgroups and subalgebras thereof.

If the symmetry is not explicitly broken under quantization, then we can
conclude that the quantized gauge fields associated with the Lie
algebra $\mathcal{G}_i$ describe gauge bosons characterized by the
set of real roots $\Bold{q}_i$. We will refer to these as \emph{kinematical quantum numbers} for the gauge bosons. They correspond
to conserved properties of the gauge bosons for
unbroken symmetries, and they are physically
relevant (though not necessarily observable) once a choice of matter field representations has been
made (in the sense that the gauge bosons and matter fields are imagined to exchange an actual `charge').

Evidently, gauge bosons associated with the $\bs{h}_{s_{i}}$ have vanishing kinematical quantum numbers while those associated with the
$\bs{e}_{\alpha_{k_i}}$ carry the $\Bold{q}_{i}$ kinematical
quantum numbers. Note that $\bs{e}_{-\alpha_{k_i}}$ carries
$-\Bold{q}_{i}$. It is in this sense that the Lie algebra
basis, defined by the (properly restricted) decomposition
(\ref{decomposition}), characterizes the physical gauge bosons.

Turn now to the matter fields. We will confine our
attention to Dirac spinors. (The
spinor components of the matter fields will not be displayed since we work in Minkowski
space-time and they play no role here in internal
symmetries.) Let $V_{\sBold{R}_i}$ be a vector space with $\mathrm{dim}V_{\sBold{R}_i}=:d_{R_i}$ that
furnishes a representation of $G_i$ having basis
$\{\Bold{e}_{l_i}^{(\sBold{R}_i)}\}_{l_i=1}^{d_{R_i}}$. And let
$\rho^{(\sBold{R}_i)}:G_i\rightarrow GL(V_{\sBold{R}_i})$ denote a
faithful irrep. The $\Bold{R}_i$ is a collection
$(R_i^1,\ldots,R_i^{d_{0_i}})$ of $d_{0_i}$ numbers and serves to
label the representation. (Recall $d_{0_i}=\mathrm{dim}V_{0_i}$).

Given some set $\{\Bold{R}_i\}$, suppose the corresponding set of
fields $\{\Bold{\Psi}^{(\sBold{R}_i)}\}$ furnish inequivalent
irreps $\rho^{(\sBold{R}_i)}(G_i)$ of the $G_i$. The associated
tensor product representation
\begin{equation*}
\rho^{(\times \sBold{R}_n)}(\times_n G_i
):=\rho^{(\sBold{R}_1)}(G_1)\times,\ldots,\times\rho^{(\sBold{R}_n)}(G_n)
\end{equation*}
of the direct product group is also irreducible (where $\times
\Bold{R}_n:=(\Bold{R}_1,\ldots,\Bold{R}_n)$ denotes an element in
the cartesian product
$\{\Bold{R}_1\}\times,\ldots,\times\{\Bold{R}_n\}$). In fact for
the class of groups under consideration, all irreps of $G$
are comprised of \emph{all possible combinations} of relevant
$\{\Bold{R}_i\}$ \cite{COR}. The corresponding Lie algebra representation
\begin{equation*}
\rho_e{'{^{(\times \sBold{R}_n)}}}(\oplus_n
\mathcal{G}_i):={\rho_{e}{'}{^{(\sBold{R}_1)}}}(\mathcal{G}_1)
\oplus,\ldots,\oplus{\rho_{e}{'}{^{(\sBold{R}_n)}}}(\mathcal{G}_n)
\end{equation*}
(where $\rho_e'$ is the derivative map of the representation
evaluated at the identity element) is likewise irreducible for all
combinations of $\{\Bold{R}_i\}$ that are associated with irreps
of the $\mathcal{G}_i$.

Our supposition is that these representations have the potential to be realized in a physical system and so \emph{all combinations should be included in realistic models}. The idea, of course, is these relevant combinations of irreps can be identified with elementary fields.

The representations $\rho^{(\sBold{R}_i)}(G_i)$ are largely a
matter of choice depending on physical input. By assumption, the
internal degrees of freedom associated with $G_i$ of
\emph{elementary} particles correspond to the basis elements
$\{\Bold{e}_{l_i}^{(\sBold{R}_i)}\}_{l_i=1}^{d_{R_i}}$ spanning
$V_{\sBold{R}_i}$. Hence, a given label $\Bold{R}_i$
characterizes the elementary particles (along with Lorentz
labels). In particular, a basis is chosen such that the
representation of the diagonal Lie algebra elements is (no
summation implied)
\begin{equation}\label{full representation}
\rho_{e}{'}{^{(\sBold{R}_i)}}(\bs{h}_{s_i})\Bold{e}_{l_i}^{(\sBold{R}_i)}
=i q^{(m)}_{s_i,l_i}\Bold{e}_{l_i}^{(\sBold{R}_i)}
\end{equation}
where $q^{(m)}_{s_i,l_i}$ are $(d_{0_i}\times d_{R_i})$  real numbers and the $(m)$ superscript indicates
``matter''. In an obvious short-hand notation,
\begin{equation}\label{full representation 2}
  \rho_{e}{'}{^{(\sBold{R}_i)}}(\bs{h}_{s_i})\Bold{e}^{(\sBold{R}_i)}
=i\Bold{q}^{(m)}_{s_i}\Bold{e}^{(\sBold{R}_i)}\,.
\end{equation}
where
$\Bold{q}^{(m)}_{s_i}:=(q^{(m)}_{s_i,1},\ldots,q^{(m)}_{s_i,d_{R_i}})$
and $\Bold{e}^{(\sBold{R}_i)}
=(\Bold{e}_1^{(\sBold{R}_i)},\ldots,\Bold{e}_{d_{R_i}}^{(\sBold{R}_i)})$ with no implied summation.
Hence, $\Bold{q}^{(m)}_{s_i}$ can serve to label the basis
elements corresponding to elementary matter particle states for a
given representation labelled by $\Bold{R}_i$. In this sense, the
elementary matter particles carry the \emph{kinematical
quantum numbers} $\Bold{q}^{(m)}_{s_i}$.

Taking the complex conjugate of (\ref{full representation}), gives
\begin{equation}\label{anti-representation 1}
[\rho_{e}{'}{^{(\sBold{R}_i)}}(\bs{h}_{s_i})]^{*}
{\Bold{e}_{l_i}^{(\sBold{R}_i)}}^{*}
=-i q^{(m)}_{s_i,l_i}{\Bold{e}_{l_i}^{(\sBold{R}_i)}}^{*}\;.
\end{equation}
Hence,
\begin{equation}\label{anti-representation 2}
{\Bold{e}^{(\sBold{R}_i)}}^{\dag}
[\rho_{e}{'}{^{(\sBold{R}_i)}}(\bs{h}_{s_i})]^{\dag}
=-i\Bold{q}^{(m)}_{s_i}{\Bold{e}^{(\sBold{R}_i)}}^{\dag}\;.
\end{equation}
So $\{{\Bold{e}_{l_i}^{(\sBold{R}_i)}}^{\dag}\}$ furnishes a dual complex-conjugate representation of $G_i$ and is obverse to
$\{{\Bold{e}_{l_i}^{(\sBold{R}_i)}}\}$. That is,
$\{{\Bold{e}_{l_i}^{(\sBold{R}_i)}}^{\dag}\}$ represents the
internal degrees of freedom of the anti-$G_i$-particles associated
with $\{{\Bold{e}_{l_i}^{(\sBold{R}_i)}}\}$ since they are
characterized by opposite quantum numbers.

The analysis in this subsection has yielded two insights that may
be useful in model building. First, the Lie algebra possesses a
distinguished basis, the Cartan basis, that is particularly suited to model gauge bosons
and their physical attributes. Second, the matter field irreps for
the direct product group $G=\times_n G_i$ include all
combinations of irreps of the subgroups $G_i$. Therefore (and we want to emphasize this) all irrep combinations should be included in model Lagrangians, and this leads to the physical realization of elementary particles possessing all combinations of kinematical quantum numbers.

Of course kinematics is not the whole story, and we must somehow relate these kinematical quantum numbers to what is observed during quantum dynamics.

\subsection{Dynamical quantum numbers}\label{sec. dynamics}
Again, the setup for gauge field theory is well-known, but we quickly review it here to set notation for a product symmetry group $G=\times_n G_i$.

Consider a principal fiber bundle with structure group $G_i$ and
Minkowski space-time base space. Let
$\mathcal{A}_i(x):=\Bold{A}^{a_i}(x)\otimes\bs{g}_{a_i}$ be the
local coordinate expression on the base space of the gauge
potential (the pull-back under a local trivialization of the
connection defined on the principal bundle). $\Bold{A}^{a_i}(x)$
is a real one-form on the base space whose
components $A_{\mu}^{a_i}(x)$ represent gauge fields. The gauge
field self-interactions are encoded in the covariant derivative of
the gauge potential
\begin{equation}\label{field strength}
  \mathcal{F}_i(x):=D\mathcal{A}_i(x)=d\mathcal{A}_i(x)
  +\tfrac{1}{2}[\mathcal{A}_i(x),\mathcal{A}_i(x)]
  =:\Bold{F}^{a_i}(x)\bs{g}_{a_i}
\end{equation}
where $\Bold{F}^{a_i}$ is a two-form on the base space. In the
special basis determined by the decomposition of the previous
section, the commutator term describes interactions between gauge
fields characterized by the kinematical quantum numbers
$\Bold{q}_i$ by virtue of (\ref{eigenvalues}).

Matter fields will be sections of a tensor product bundle
$S\otimes V$. Here $S$ is a spinor bundle over space-time with
typical fiber $\C^4$, and $V$ is a vector bundle associated to the
gauge principal bundle with typical fiber
$V_{\times\sBold{R}_n}:=\otimes V_{\sBold{R}_i}$.

A basis element in $\C^4\otimes V_{\times\sBold{R}_n}$ will be
denoted $\Bold{e}_{\times l_n}^{(\times\sBold{R}_n)}
:=\otimes\Bold{e}_{l_i}^{(\sBold{R}_i)}$. (For clarity, we will
not make the spinor index explicit.) Vector space
$V_{\times\sBold{R}_n}$ furnishes the representation
$\rho{{^{(\times \sBold{R}_n)}}}(\times_n G_i)$. It is this
representation that determines the gauge--matter field interactions
via the covariant derivative $\D$;
\begin{equation}\label{matter field covariant derivative}
  \D\Bold{\Psi}^{(\times\sBold{R}_n)}(x)=\left[\not\!{\partial}
  +\rho_e{'{^{(\times \sBold{R}_n)}}}(\not\!\!\mathcal{A})
  \right]\Bold{\Psi}^{(\times\sBold{R}_n)}(x)
\end{equation}
where $\not\!\!\mathcal{A}:=
\mathrm{i}_{\gamma}\mathcal{A}=\gamma_\mu\mathcal{A}^\mu\in\oplus_n\mathcal{G}_i$ and
$\Bold{\Psi}^{(\times\sBold{R}_n)}(x):=\mathit{\Psi}^{\times
l_n}(x)\Bold{e}_{\times l_n}^{(\times\sBold{R}_n)}$.

There is a scale ambiguity that resides in the matter field
covariant derivative. The inner product on
$\rho_e'(\mathcal{G}_i)$ for any \emph{faithful} representation is proportional to the inner
product on $\mathcal{G}_i$. This implies the matrices in the
covariant derivative (\ref{matter field covariant derivative}) are
determined only up to overall constants $\kappa_{\mathcal{G}_i}$
--- relative to the scale of the gauge fields.  These constants are
conventionally interpreted as coupling constants characterizing
the gauge boson--matter field interaction. We choose the coupling
constants so that, given gauge and matter field normalizations,
the parameters in the matter field covariant derivative that
characterize \emph{neutral} gauge--matter field interactions coincide with
the matter field kinematical quantum numbers $\Bold{q}_{s_i}^{(m)}$. With this choice, \emph{the parameters characterizing couplings in
both the gauge and matter field covariant derivatives are proportional to the
kinematical quantum numbers associated with the weights of the associated representation}.

Now, the (bare) Lagrangian density that determines the dynamics is comprised of the usual Yang-Mills
terms, spinor matter field terms, ghost terms, and gauge fixing
terms. The Yang-Mills terms are
\begin{equation}\label{Yang-Mills}
  -\frac{1}{2}\sum_i\mathcal{F}_i\cdot\mathcal{F}_i
\end{equation}
where the dot product
represents both the Minkowski metric and an $Ad(g_i)$ invariant inner
product on each $\mathcal{G}_i$. For reductive $G_i$, the $Ad$-invariant inner product on each subspace determined by  $\mathrm{span}_{\C}\{\bs{g}_{a_i}\}$ is
classified by two real constants.\footnote{For any two elements $\bs{g}_{a_i}$ and $\bs{g}_{b_i}$ there are two independent trace combinations, viz. $\mathrm{tr}(\bs{g}_{a_i}\bs{g}^\dag_{b_i})$ and $\mathrm{tr}(\bs{g}_{a_i})\mathrm{tr}(\bs{g}^\dag_{b_i})$. So a general Hermitian bilinear form is a real linear combination of these two. Of course for $SU(N)$ the second combination vanishes identically, and the inner product is then classified by a single constant.} The normalization chosen for the inner product effectively fixes the scale of
$\Bold{A}^{a_i}(x)$ and hence also the gauge fields
$A_{\mu}^{a_i}(x)$ given the standard Minkowski inner product.

The most general (quadratic, minimal coupling) matter field  Lagrangian kinetic term consistent
with the requisite symmetries is, according to the suggestion from
the previous section, comprised of a sum over all the
faithful irreps (including complex conjugates) of the elementary matter fields:
\begin{equation}\label{matter lagrangian}
  \mathcal{L}_m
  =i\sum_{\times\sBold{R}_n}\kappa_{\times\sBold{R}_n}
  \overline{\Bold{\Psi}}^{(\times\sBold{R}_n)}\cdot
  \D\Bold{\Psi}^{(\times\sBold{R}_n)}+\mbox{mass terms}
\end{equation}
where $\overline{\Bold{\Psi}}^{(\times\sBold{R}_n)}$ is a section
of the dual complex-conjugate bundle $\overline{S\otimes V}=\overline{S}\otimes
\overline{V}$ and $\kappa_{\times\sBold{R}_n}$ are positive real
constants that are constrained by various consistency conditions;
for example, anomaly considerations and CPT symmetry. It is clear
that $\delta\mathcal{L}_m=0$ for $\Bold{\Psi}(x)\rightarrow
\exp\{\theta(x)^{a_i}\rho_e{'}(\bs{g}_{a_i})\}\Bold{\Psi}(x)$
despite the presence of $\kappa_{\times\sBold{R}_n}$ (assuming
appropriate mass terms).

The dot product here represents a Lorentz and $\rho(g)$ invariant
Hermitian matter field inner product; a bundle metric. Recall the matter fields furnish complex representations. For non-isomorphic representations, the scaling constants $\kappa_{\times\sBold{R}_n}$ can be absorbed into the inner product on the representation space. However, for the special case of representations that are related via the adjoint action of a \emph{unitary} symmetry group, their underlying vector spaces are isomorphic. Consequently their inner products are tied together and the associated ratios of $\kappa_{\times\sBold{R}_n}$ can be non-trivial. This
persists even after renormalization. The
possibility of non-trivial factors $\kappa_{\times\sBold{R}_n}$ in
the matter field Lagrangian density will be a key element in our
non-Standard Model.

For each individual subgroup $G_i$, the gauge and matter field
terms in the Lagrangian density give rise to the conserved
currents
\begin{equation}\label{conserved current 1}
J_{(a_i)}^{\mu}=
-\mathcal{F}_i^{\mu\nu}\cdot\left[\bs{g}_{a_i},{\mathcal{A}_i}_{\nu}\right]
  +j_{(a_i)}^{\mu}
\end{equation}
where
\begin{equation}\label{conserved current 2}
  j_{(a_i)}^{\mu}=\sum_{\times\sBold{R}_n}
  \kappa_{\times\sBold{R}_n}
  \overline{\Bold{\Psi}}^{(\times\sBold{R}_n)}
  \cdot\gamma^{\mu}\rho_e{'{^{(\times\sBold{R}_n)}}}
  (\bs{g}_{a_{i}})\Bold{\Psi}^{(\times\sBold{R}_n)}
\end{equation}
are the covariantly conserved matter field currents. In
particular, the neutral conserved currents  associated with $G_i$
are
\begin{eqnarray}\label{conserved current 3}
  J_{(s_i)}^{\mu}&=&
-\mathcal{F}_i^{\mu\nu}\cdot\left[\bs{h}_{s_i},
{\mathcal{A}_i}_{\nu}\right]
  +j_{(s_i)}^{\mu}\notag\\
  &=&-q_{s_i}F_{-\alpha_{k_i}}^{\mu\nu}A^{\alpha_{k_i}}_{\nu}
  +\sum_{\times\sBold{R}_n}\kappa_{\times\sBold{R}_n}
  (\Bold{q}^{(m)}_{s_i})
  \overline{\Bold{\Psi}}^{(\times\sBold{R}_n)}
  \cdot\gamma^{\mu}\Bold{\Psi}^{(\times\sBold{R}_n)}\,.
\end{eqnarray}
The constants $\kappa_{\times\sBold{R}_n}(\Bold{q}^{(m)}_{s_i})$
will be termed `coupling strengths', and they represent the
scale of gauge--matter field couplings given matter field
normalizations. Evidently, not all matter field currents
contribute to interactions on an equal basis if $\kappa_{\times\sBold{R}_n}$ is non-trivial.

This is significant
because some particles characterized by a set of kinematical quantum
numbers may appear to have scaled coupling strengths  when
interacting with gauge bosons. However, in order to conclude this, we must first confirm that the
normalization freedom in the Lagrangian density allows us to
maintain equality between the \emph{renormalized} parameters
${q}_{s_i}$ and $\Bold{q}^{(m)}_{s_i}$ appearing in the quantum field relations expressed below in equation
(\ref{charge operators}) and the kinematical quantum numbers discussed in the previous section.
Moreover, we must verify that non-vanishing scaling parameters $\kappa_{\times\sBold{R}_n}$ do
not destroy the assumed local symmetries.

To that end, consider the neutral quantum charge operators $Q_{(s_i)}:=-i\int
\hat{J^0}_{(s_i)}dV$ associated with the currents given by (\ref{conserved current 3}). They encode \emph{dynamical quantum numbers} in the sense that
\begin{eqnarray}\label{charge operators}
  &&[Q_{(s_i)},A^{\alpha_{k_j}}_{\perp}]
  =q_{s_i}A^{\alpha_{k_j}}_{\perp}\delta_{ij}\notag\\
   &&[Q_{(s_i)},\Bold{\Psi}^{(\times\sBold{R}_n)}]
   =\Bold{q}^{(m)}_{s_i}\Bold{\Psi}^{(\times\sBold{R}_n)}
\end{eqnarray}
where the gauge and matter fields have been promoted to quantum
operators and $A^{\alpha_{k_i}}_{\perp}$ are the transverse gauge
fields. The second relation follows because the conjugate momentum
of $\Bold{\Psi}^{(\times\sBold{R}_n)}$ is
$\kappa_{\times\sBold{R}_n}\overline{\Bold{\Psi}}^{(\times\sBold{R}_n)}\gamma^0$
as determined from (\ref{matter lagrangian}).

Equations (\ref{charge operators}) are in terms of bare
quantities, but they are required to be valid for renormalized quantities
as well. Under the renormalizations
\begin{equation}\label{renormalization 1}
\mathcal{A}_{i}^{\mathrm{B}}\rightarrow
Z_{\mathcal{A}_{i}}^{1/2}\mathcal{A}_{i}^{\mathrm{R}}
\end{equation}
and
\begin{equation}\label{renormalization 2}
{\Bold{\Psi}^{(\times\sBold{R}_n)}}^{\mathrm{B}}\rightarrow
Z_{\sBold{\Psi}^{(\times\ssBold{R}_n)}}^{1/2}{\Bold{\Psi}^{(\times\sBold{R}_n)}}^{\mathrm{R}}\;,
\end{equation}
the basis elements $\bs{g}_{a_i}$
can be re-scaled so that
$\Bold{q}_i^{\mathrm{B}}=Z_{\mathcal{A}_i}^{-1/2}\Bold{q}_i^{\mathrm{R}}$.
Likewise, the basis $\Bold{e}_{l_i}^{(\sBold{R}_i)}$ can be
re-scaled so that ${\Bold{q}_{s_i}^{(m)}}^{\mathrm{B}}
=Z_{\sBold{\Psi}^{(\times\ssBold{R}_n)}}^{-1/2}{\Bold{q}_{s_i}^{(m)}}^{\mathrm{R}}$.

Consequently the relations (\ref{charge operators}) will be
maintained under renormalization. The renormalized form of
equations (\ref{charge operators}) are to be compared to
(\ref{eigenvalue}) and (\ref{full representation}). That they are
consistent is a consequence of: i) the covariant derivatives
(\ref{field strength}) and (\ref{matter field covariant
derivative}), ii) our choice of Lie algebra inner product, and
iii) identifying the renormalized dynamical quantum
numbers with the kinematical quantum
numbers. This consistency ensures
the renormalized gauge and matter fields appearing in the
Lagrangian density can be identified with the elementary fields associated with the Lie algebra-induced
quantum numbers $\Bold{q}_{i}$ and $\Bold{q}_{s_i}^{(m)}$. It
should be emphasized that the gauge group coupling constants are implicit in
$\Bold{q}_{i}$ and $\Bold{q}_{s_i}^{(m)}$, and non-trivial
$\kappa_{\times\sBold{R}_n}$ do not get renormalized; or, rather,
non-trivial $\kappa_{\times\sBold{R}_n}$ persist after
renormalization of $\Bold{\Psi}^{(\times\sBold{R}_n)}$.

Now, to maintain the all-important local symmetries of the Lagrangian
density, $Q_{(a_i)}$ and $\{\bs{g}_{a_i}\}$, along with their
associated commutation relations, must determine isometric
algebras. Fortunately, we find
\begin{eqnarray}\label{current algebra}
  [\hat{J^0}_{(a_i)},\hat{J^0}_{(b_j)}]&=&\delta_{ij}C^{c_j}_{a_ib_j}
\left\{-\mathcal{F}_i^{\mu\nu}\cdot\left[\bs{g}_{c_j},
{\mathcal{A}_i}_{\nu}\right]
 +\sum_{\times\sBold{R}_n}
  \kappa_{\times\sBold{R}_n}
  \overline{\Bold{\Psi}}^{(\times\sBold{R}_n)}
  \rho_e{'{^{(\times\sBold{R}_n)}}}
  (\bs{g}_{c_j})\Bold{\Psi}^{(\times\sBold{R}_n)}\right\}\notag\\\notag\\
  &= & \delta_{ij}C^{c_j}_{a_ib_j}\hat{J^0}_{(c_j)}
\end{eqnarray}
where $C_{a_ib_i}^{c_i}$ are the structure constants of
$\mathcal{G}_i$.

It is crucial that the $\kappa_{\times\sBold{R}_n}$ factors
do not spoil the equality between the kinematical and dynamical quantum numbers or the local symmetries.
Given these developments, it makes sense to refer to the two types of quantum
numbers --- renormalized dynamical quantum numbers and kinematical
quantum numbers --- by the common term \emph{intrinsic charges}. On the other hand, the
renormalized coupling strengths
$\kappa_{\times\sBold{R}_n}({\Bold{q}_{s_i}^{(m)}}^\mathrm{R})$ in
the renormalized currents (\ref{conserved current 3}) will be
called \emph{extrinsic charges}.

The analysis in this subsection leads to the conclusion that,
\emph{in some cases, the intrinsic charges of matter
fields do not fully determine their coupling strengths to gauge
bosons}. Stated otherwise, the intrinsic and extrinsic charges of quantized matter fields are not necessary equivalent. (Evidently, the qualifier intrinsic/extrinsic is not necessary for gauge bosons.)

\section{The non-Standard Model}
Instead of the SM symmetry group $SU_C(3)\times SU_I(2)\times U_Y(1)$ based on color, isospin, and hypercharge; we choose the non-Standard Model (n-SM) symmetry group $SU_S(3)\times U_{EW}(2)$ with associated Lie
algebra $\mathfrak{su}_S(3)\oplus \mathfrak{u}_{EW}(2)\cong \mathfrak{su}_S(3)\oplus \mathfrak{u}_{EW}(2)/\mathfrak{u}_{EM}(1)\oplus \mathfrak{u}_{EM}(1)$ based on strong charge (as opposed to color charge as will be explained below) and electric charge.

\subsection{Gauge boson charges}
In our picture, gauge boson charges are determined by the Cartan basis.

The root system for $SU(3)$ is two-dimensional implying two strong charges. Hence the Cartan decomposition yields two strong-charge-neutral gluons and three sets of oppositely strong-charged gluons. Accordingly, our interpretation of strong charge associated with $SU_{S}(3)$ differs from the standard assignment of color charge, because we associate charge with roots (quantum numbers) rather than degrees of freedom in the defining representation.\footnote{Hence, we write $SU_{S}(3)$ instead of $SU_{C}(3)$ to emphasize this interpretational difference, and $U_{EW}(2)$ instead of $U_{(I,Y)}(2)$ to emphasize our notion of `physical' electroweak gauge bosons that carry electric and weak charge as opposed to isospin and hypercharge.} For book-keeping purposes, there is no essential difference between the two interpretations \emph{in the defining representation} since there is a one-to-one correspondence between color degrees of freedom and matter field eigenstates with non-vanishing strong charge. (In other words, matter field eigenstates are kinematically gauge equivalent.) However, this is not the case for the adjoint representation where the interaction landscape differs because the Cartan subalgebra is invariant under inner automorphisms: Neutral and strong-charged gluons can interact but they don't mix under gauge transformations, and this will be reflected in the dynamics.

\begin{remark}Conventionally, referring to the tensor product of the fundamental representation,
gluons are thought to carry one color charge and one anti-color charge, and they exchange one of these color charges in a quark/gluon interaction. But there is nothing forcing the conserved charges that characterize gluon interactions to align with ``color'' degrees of freedom. For us, color simply labels the degrees of freedom in the defining representation and has nothing to do with strong charge --- although it still plays a role in classifying irreducible representations in the usual way.

Our interpretation of strong charge differs
substantially: Indeed, from our standpoint strong charges, being based on the Lie algebra root system\footnote{Of course, QCD practitioners are well aware of Lie algebra root systems, but none of the pedagogical treatments of QCD (that we are aware of) identify the roots with gluon charges.}, are hierarchical (in the sense of non-trivial charge ratios). This hierarchy will induce some self-energy differences among gluons since neutral gluons don't directly interact.

The two neutral
gluons, which apparently experience a quite different confining potential compared to charged gluons, have obvious implications regarding nuclear forces. Specifically, one can imagine the neutral gluons (at least partly) mediating the nuclear force.\footnote{However, being traceless, neutral gluons do not manifest as long-range force carriers between strong-singlet states.}
\end{remark}

Moving on to the $U_{EW}(2)$ subgroup, there will be two neutral gauge bosons and a pair of oppositely charged gauge bosons carrying two types of electroweak charge (electric and weak). However, in the broken symmetry sector, the only gauge boson that survives is a neutral boson that characterizes electric charge (and the mass and electric charge that now distinguish the broken symmetry generators stand in for their original electroweak charges). There is no compelling reason to introduce isospin and hypercharge since experiment dictates that the charge associated with the unbroken gauge boson is what we know as electric charge. Of course, in the unbroken symmetry phase, the Cartan basis characterizing electric charge is only unique modulo $U_{EW}(2)$ conjugation. Consequently, one can talk about any other gauge-equivalent charge combination consistent with the relations imposed by the Lie algebra decomposition described in $\S\,2$. Note, however, that isospin and hypercharge are not gauge equivalent to the two EW charges since the hypercharge generator commutes with every other generator.

\subsection{Hadronic Constituents} \label{hadronic constituents}

We will consider only Dirac matter fields in the fundamental
representation of $SU(3)$ and $U(2)$. Consequently, the matter
fields are sections of an associated fiber bundle with typical
fiber $\C^4\otimes\C^3\otimes\C^2$. Since $\mathfrak{su}(3)$ and $\mathfrak{u}(2)$ are
both rank two algebras, these matter fields can be labelled by
four quantum numbers; two associated with $SU(3)$ and two
with $U(2)$. According to the previous section, the relevant irreps of the direct product
group are postulated to include the $(\bs{3},\bs{2})$ and
$(\bs{3},{\mathbf{2^c}})$ and their anti-fields
$(\overline{\bs{3}},\overline{\bs{2}})$ and
$(\overline{\bs{3}},\overline{{\mathbf{2^c}}})$.

Spinor fields in the defining representation require the product bundle $S\otimes
V_{\sBold{R}}$ where $S$ is a spinor bundle over Minkowski
space-time. For example, given a trivialization of the bundle $S\otimes
V_{(\mathbf{3},\mathbf2)}$, let $\{\Bold{e}^{\alpha
Aa}\}:=\{\Bold{\psi}^{\alpha}\otimes\Bold{e}^A\otimes\Bold{e}^a\}$
be the chosen basis that spans the typical fiber
$\C^4\otimes\C^{\,\text{3}}\otimes\C^{\,\text{2}}$. (Indices are
assumed to have the necessary ranges for any given
representation.) Sections $\Bold{\Psi}=\mathit{\Psi}_{\alpha
Aa}\Bold{e}^{\alpha Aa}$ of $S\otimes V_{(\mathbf{3},\mathbf2)}$
constitute the elementary spinor fields in the
$(\mathbf{3},\mathbf2)$ representation, and
$\Bold{e}^A\otimes\Bold{e}^a$ encode the internal
$SU(3)\times U(2)$ degrees of freedom.

For the $U_{EW}(2)$-conjugate
representation, define $\rho^{\bs{c}}:=(\tau_1)\rho^\ast(\tau_1)^{-1}$ and ${\Bold{\Psi}^{\bs{c}}}:=(\tau_1)\Bold{\Psi}^\ast$ where
$\tau_1=\left(
\begin{array}{cc}
0 & 1 \\
1 & 0 \\
\end{array}
\right)\in U(2)$. Explicitly,  ${\Bold{\Psi}^{\bs{c}}}$
is a section of the bundle $S\otimes V_{SU(3)}\otimes
{V}^{\bs{c}}_{U(2)}$ which is the image under the bundle morphism
\begin{eqnarray}
F:S\otimes V_{SU(3)}\otimes V_{U(2)}&\longrightarrow& S\otimes
V_{SU(3)}\otimes {V}^{\bs{c}}_{U(2)}\notag\\
\rule{0in}{.2in}
(x,\mathit{\Psi}_{\alpha Aa}(x)\Bold{e}^{\alpha Aa})&\longmapsto&
(x,[\tau_1]^{b}_a\mathit{\Psi}^\ast_{\alpha Ab}(x)\Bold{e}^{\alpha Aa})
=:(x,{\mathit{\Psi}}^{\bs{c}}_{\alpha Aa}(x)(\Bold{\psi}^{\alpha}\otimes\Bold{e}^A\otimes
\Bold{e}_{\bs{c}}^{a}))\notag\\
\end{eqnarray}
in a given chart and trivialization.  Note that $\{\Bold{e}_{\bs{c}}^{a}\}^{\bs{c}}=\{\Bold{e}^{a}\}$. Meanwhile, the dual complex-conjugate representation is given by ${\rho}^\dag=(\tau_1)^2{{\rho}^\dag}(\tau_1)^{-2}
=(\tau_1){\rho^{\bs{c}}}^{\mathrm{T}}(\tau_1)^{-1}$ and
\begin{eqnarray}\label{conjugate}
i\Bold{q}^{(m)}_{s^{\bs{c}}}\Bold{\Psi}^{\bs{c}}
=\rho^{\bs{c}}_e{'}(\bs{h}_s)\Bold{\Psi}^{\bs{c}}
 =(\tau_1)(\rho_e{'}(\bs{h}_s)\Bold{\Psi})^\ast
 =(\tau_1)(-i\Bold{q}^{(m)}_{s}\Bold{\Psi}^\ast)
 =-i\Bold{q}^{(m)}_{s}\Bold{\Psi}^{\bs{c}}\;.
\end{eqnarray}
Hence, interpret $\Bold{\Psi}^{\bs{c}}$  as the `` $U(2)$ anti-field'' of $\Bold{\Psi}$ in the sense that $U_{EW}(2)$-conjugation yields a field with opposite $U_{EW}(2)$ charges with respect to a permuted basis. (We emphasize that there is no conjugation associated with the $SU(3)$ or spinor index here.)

The covariant derivatives acting on the matter fields in the
$(\bs{3},\bs{2})$ and $(\bs{3},{\mathbf{2^c}})$ representations
are
\begin{equation}\label{covariant derivative 1}
  (\D\Bold{\Psi})=\left\{\not\!{\partial}[\bs{1}]_{Aa}^{Bb}
  +\not\!G^{\alpha}\,[\bs{\Lambda}_{\alpha}]_A^B\otimes [\bs{1}]_{a}^{b}
  +[\bs{1}]_{A}^{B}\;\otimes\not \!g^{\sigma}\, [\Bold{\lambda}_{\sigma}]_a^b\;\right\}
  \mathit{\Psi}_{Bb}\Bold{e}^{Aa}
\end{equation}
and
\begin{equation}\label{covariant derivative 2}
  ({\D}^{\bs{c}}{\Bold{\Psi}^{\bs{c}}})=\left\{\not\!{\partial}[\bs{1}]_{Aa}^{B{b}}
  +\not\!G^{\alpha}\,[\bs{\Lambda}_{\alpha}]_A^B\otimes [\bs{1}]_{a}^{{b}}
  +[\bs{1}]_{A}^{B}\;\otimes\not {\!{g^{\sigma}}}^\ast\, [{\Bold{\lambda}^{\bs{c}}}_{\sigma}]_{a}^{{b}}
  \;\right\}
  \mathit{\Psi}^{\bs{c}}_{B{b}}\Bold{e}^{Aa}_{\bs{c}}
\end{equation}
respectively with $\Bold{\lambda}_{\sigma}$ and $\bs{\Lambda}_{\alpha}$ generating $U_{EW}(2)$ and $SU_{C}(3)$. These yield the kinetic matter field Lagrangian density;
\begin{eqnarray}\label{matter field Lagrangian}
  \mathcal{L}_m&=&i\kappa\overline{\mathit{\Psi}_{A'a'}}\left\{\not\!{\partial}[\bs{1}]_{Aa}^{Bb}
  +\not\!G^{\alpha}\,[\bs{\Lambda}_{\alpha}]_A^B\otimes [\bs{1}]_{a}^{b}
  +\not \!g^{\sigma}\,[\bs{1}]_{A}^{B}\otimes [\Bold{\lambda}_{\sigma}]_a^b\;\right\}
  \mathit{\Psi}_{Bb}\delta^{A'A}\delta^{a'a}\notag\\
  &&+i\kappa^{\bs{c}}\overline{\mathit{\Psi}^{\bs{c}}_{A'{a}'}}
  \left\{\not\!{\partial}[\bs{1}]_{Aa}^{B{b}}
  +\not\!G^{\alpha}\,[\bs{\Lambda}_{\alpha}]_A^B\otimes [\bs{1}]_{a}^{{b}}
  +\not\!{g^{\sigma}}^\ast\,[\bs{1}]_{A}^{B}\otimes [{\Bold{\lambda}}^{\bs{c}}_{\sigma}]_{a}^{{b}}
  \;\right\}
  \mathit{\Psi}^{\bs{c}}_{B{b}}\delta^{A'A}\delta^{{a}'a}\;
\end{eqnarray}
where $\kappa,\kappa^{\bs{c}}$ are (for now) undetermined constants. Note that it is not possible to absorb both $\kappa$ and
$\kappa^{\bs{c}}$ by separate field redefinitions because
$\mathit{\Psi}^{\bs{c}}_{Aa}=[\tau_1]_a^{b}\mathit{\Psi}^\ast_{Ab}$
and $\Bold{e}^a\cdot\Bold{e}^a=\Bold{e}^{a}_{\bs{c}}\cdot\Bold{e}^{a}_{\bs{c}}$;
implying that the ratio $\kappa^{\bs{c}}/\kappa$ can be non-trivial
in this example.

The corresponding $U(2)$ and $SU(3)$ currents are
\begin{equation}\label{currents 1}
  j^{\mu}_{(\sigma)}=\kappa\overline{\mathit{\Psi}_a^{A}}\gamma^{\mu}
  [\Bold{\lambda}_{\sigma}]^a_b
  \mathit{\Psi}_A^{b}
  +\kappa^{\bs{c}}\overline{{\mathit{\Psi}^{\bs{c}}}^A_{a}}
  \gamma^{\mu}[{\Bold{\lambda}^{\bs{c}}}_{\sigma}]^{a}_{{b}}
  \mathit{\Psi}_A^{{b}}
\end{equation}
and
\begin{equation}\label{currents 2}
  j^{\mu}_{(\alpha)}=\kappa\overline{\mathit{\Psi}_A^{a}}\gamma^{\mu}
  [\bs{\Lambda}_{\alpha}]^A_B
  \mathit{\Psi}_a^{B}
  +\kappa^{\bs{c}}\overline{{\mathit{\Psi}^{\bs{c}}}_A^{a}}
  \gamma^{\mu}[\bs{\Lambda}_{\alpha}]^{A}_{B}{\mathit{\Psi}^{\bs{c}}}^B_{a}
  =(\kappa+\kappa^{\bs{c}})\overline{\mathit{\Psi}_A^{a}}
  \gamma^{\mu}[\bs{\Lambda}_{\alpha}]^{A}_{B}\mathit{\Psi}_a^{B}
\end{equation}
respectively. Evidently, if $(\kappa+\kappa^{\bs{c}})=1$, the
original $SU(3)$ coupling strength is preserved; i.e., the $SU(3)$
intrinsic and extrinsic charges are equivalent. In this
case, \emph{the $U(2)$ extrinsic charges are fractional relative to the
intrinsic charges} since $\kappa,\kappa^{\bs{c}}\neq 0$ by
assumption. In a viable model, as we will see presently, the ratio is ultimately fixed
by anomaly considerations.

The structure of these currents suggests to define hadronic constituents (HC) in the
$(\mathbf{3},\mathbf2)$ representation by $\Bold{H}^{+}:=H^+_A\Bold{e}^A$ and
their $\newew$-conjugates $(\mathbf{3},{\mathbf2}^{\bs{c}})$ by
$\Bold{H}^-:={H^-_A}{\Bold{e}^A}$ where
\begin{eqnarray} \label{matter fields}
H^{+}_{A}:=\mathit{\Psi}_{Aa}\Bold{e}^a
=\mathit{\Psi}_{A1}\Bold{e}^1+\mathit{\Psi}_{A2}\Bold{e}^2
=:(h^+\Bold{e}^1)_A+ (\xi^0\Bold{e}^2)_{A}
=\begin{pmatrix} h^{+}\\
\xi^{0}
\end{pmatrix}_{A}
\end{eqnarray}
and
\begin{eqnarray}\label{matter fields 2}
H^{-}_{A}:=[\tau_1]_{a}^b\mathit{\Psi}^\ast_{Ab}\Bold{e}^a
=\mathit{\Psi}^{\bs{c}}_{A1}\Bold{e}^1_{\bs{c}}+\mathit{\Psi}^{\bs{c}}_{A2}\Bold{e}_{\bs{c}}^2
=\mathit{\Psi}^\ast_{A2}\Bold{e}^1+\mathit{\Psi}^\ast_{A1}\Bold{e}^2
=\begin{pmatrix} {{\xi^{0}}^\ast}\\ {h^{+}}^\ast
\end{pmatrix}_{A}
=:\begin{pmatrix} {\chi^{0}}\\ h^{-}
\end{pmatrix}_{A}
\end{eqnarray}
Here $h^{\pm},\,\xi^0,\,\chi^0$ are complex space-time Dirac spinor
fields and
$\Bold{e}^{1,2}$ span $\C^2$. The superscripts on the component fields denote electric charge only since weak charge is not relevant in the broken sector. Being components of hadronic constituents and possessing integer electric charge, we will  give $h^{\pm},\,\xi^0,\,\chi^0$ the name `iquarks'.
There are three copies of $\Bold{H}^{\pm}$ accounting
for the three iquark generations. No generality is sacrificed by
assuming $\Bold{H}^{\pm}$ are normalized.

By assumption, both the left and right-handed iquarks furnish the
$\mathbf3$ of $SU_S(3)$. Also by assumption, the left-handed iquarks furnish the
$\mathbf2$ and ${\mathbf2}^{\bs{c}}$ of $\newew$ while the right-handed iquarks furnish
the $\mathbf1^+$, $\mathbf1^-$ and $\mathbf1^0$. Thus,
we have $\Bold{H}^+_\mathrm{L}:=1/2(1+\gamma_5)\Bold{H}^+$,
$\Bold{H}^-_\mathrm{L}$, $\Bold{h}^+_\mathrm{R}$,
$\Bold{h}^-_\mathrm{R}$, and $\Bold{\xi}^0_\mathrm{R}$ furnishing
the $(\mathbf{3},\mathbf2)$, $(\mathbf{3},{\mathbf2}^{\bs{c}})$,
$(\mathbf{3},\mathbf1^+)$, $(\mathbf{3},\mathbf1^-)$, and
$(\mathbf{3},\mathbf1^0)$ of $SU_S(3)\times U_{EW}(2)$, respectively.

Let us work out the explicit form of these representations. Start with $\newew$. Physically, in the broken symmetry regime
characterized by matter fields with conserved electric charge, the
Lie algebra $\mathfrak{u}_{EW}(2)$ decomposes according to $\mathfrak{u}_{EM}(1)\oplus
(\mathfrak{u}_{EW}(2)/\mathfrak{u}_{EM}(1))$. Consequently, the gauge bosons are also
characterized by electric charge. This implies the broken
symmetry generators are eigenvectors of the adjoint map of the
unbroken, electric charge generator. That is, the Lie algebra
decomposition is $\mathfrak{u}_{EW}(2)=\mathfrak{u}_{EM}(1)\oplus \mathrm{k}$ such that
$\mathfrak{u}_{EM}(1)\cap\mathrm{k}=0$ and
$ad(\mathfrak{u}_{EM}(1))\mathrm{k}\subseteq\mathrm{k}$.

Recall that we require, on physical grounds, a real subalgebra of $\mathfrak{u}(2,\C)$ for the gauge bosons. Since $\newew$ has
rank $2$, the relevant basis therefore obeys
\begin{eqnarray}\label{basis}
&[\Bold{\lambda}_{\pm},\Bold{\lambda}_{\mp}]&=\sum_{i}\pm  c'_i\Bold{\lambda}_i,\nonumber\\
&[\Bold{\lambda}_{\pm},\Bold{\lambda}_i]&=\pm c_i\Bold{\lambda}_{\pm},\nonumber\\
&[\Bold{\lambda}_{i},\Bold{\lambda}_j]&=0,
\end{eqnarray}
where $c_i,c'_i$ are structure coefficients and $i,j\in\{1,2\}$. The real (compact) subalgebra that generates $\newew$ is $\mathfrak{u}_{EW}(2)
:=\mathrm{span}_{\C/\sim}
\{\Bold{\lambda}_+,\Bold{\lambda}_-,\Bold{\lambda}_1,\Bold{\lambda}_2\}$ where $\C/\sim$ is the set of coefficients $(a^\pm,b_i,)$ such that $a^-=(a^+)^\ast$ with $a^\pm\in\C$ and $b_i\in\R$. Explicitly, a general skew-hermitian element $\mathfrak{g}\in\mathfrak{u}_{EW}(2)$ in this basis decomposes as $\mathfrak{g}
=a^+\Bold{\lambda}_++a^-\Bold{\lambda}_-+b_1\Bold{\lambda}_1+b_2\Bold{\lambda}_2$.

The most
general defining representation of $\mathfrak{u}_{EW}(2)$ allowed by (\ref{basis}) is
generated by
\begin{eqnarray}\label{rep}
  \bs{T}_+:=\rho'_e(\Bold{\lambda}_+)&=&
  i\begin{pmatrix}
  0 & & t \\
  0 & & 0
  \end{pmatrix},\;\;\;\;
  \bs{T}_-:=\rho'_e(\Bold{\lambda}_-)=i\begin{pmatrix}
  0 & & 0 \\
  t & & 0
  \end{pmatrix},\nonumber\\
  \bs{T}_0:=\rho'_e(\Bold{\lambda}_1)&=&i\begin{pmatrix}
  u & & 0 \\
  0 & & v
  \end{pmatrix},\;\;\;\;
  \bs{Q}:=\rho'_e(\Bold{\lambda}_2)=i\begin{pmatrix}
  r & & 0 \\
  0 & & s
  \end{pmatrix},
\end{eqnarray}
where the representation is a unitary map $\rho:\newew\rightarrow
GL(\mathbb{C}^2)$, $\rho'_e$ denotes the derivative of the
representation evaluated at the identity element
$e\in\newew$, and $r,s,t,u,v$ are  real constants.

To proceed, an $Ad$-invariant form on
$\mathfrak{u}_{EW}(2)$ is required. In fact, there is a $2$-dimensional real
vector space of $Ad$-invariant Hermitian bilinear forms on
$\mathfrak{u}_{EW}(2)$ given by (\cite{D} pg. 114)
\begin{equation}\label{bilinear}
  \langle\Bold{\lambda}_{\sigma},\Bold{\lambda}_{\rho}\rangle
  :=2g_1^{-2}\mathrm{tr}(\Bold{\lambda}_{\sigma}\Bold{\lambda}^\dag_{\rho})
  +(g_2^{-2}-g_1^{-2})\mathrm{tr}\Bold{\lambda}_{\sigma}
  \cdot\mathrm{tr}\Bold{\lambda}^\dag_{\rho}
\end{equation}
where $g_1$ and $g_2$ are real parameters and the $\dag$ operation (complex-conjugate transpose) is defined since $U(2)$ is a matrix Lie group.

Restricting to  $0<g_2^2< g_1^2$ yields a positive-definite inner product defined by
\begin{equation}\label{inner product}
  g(\Bold{\lambda}_{\sigma},\Bold{\lambda}_{\rho})
  :=-1/2\langle\Bold{\lambda}_{\sigma},\Bold{\lambda}_{\rho}\rangle\;.
\end{equation}
Explicitly, in the basis defined by (\ref{basis}),
\begin{equation}
  \left(g_{\sigma\rho}\right)=
  \begin{pmatrix}
  {g_W}^{-2} & & 0 & & 0 & & 0 \\
  0 & & {g_W}^{-2} & & 0 & & 0 \\
  0 & & 0 & & {g_Z}^{-2} & & 0 \\
  0 & & 0 & & 0 & & {g_Q}^{-2} \\
  \end{pmatrix}
\end{equation}
where
\begin{equation}
  {g_W}^{-2}:=g(\Bold{\lambda}_{\pm},\Bold{\lambda}_{\pm}),\;
  {g_Z}^{-2}:=g(\Bold{\lambda}_{1},\Bold{\lambda}_{1}),\;
  {g_Q}^{-2}:=g(\Bold{\lambda}_{2},\Bold{\lambda}_{2})\;.
\end{equation}
The inner product can be put into canonical form by re-scaling the
$\mathfrak{u}_{EW}(2)$ basis vectors by $\Bold{\lambda}_{\pm}\rightarrow
g_W\Bold{\lambda}_{\pm}$, $\Bold{\lambda}_{1}\rightarrow g_Z\Bold{\lambda}_{1}$, and
$\Bold{\lambda}_{2}\rightarrow g_Q\Bold{\lambda}_{2}$.

The inner product on the Lie algebra is
proportional to the inner product for any of its faithful
representations.  Hence, (\ref{matter fields}), (\ref{bilinear}),
(\ref{inner product}), together with the orthogonality condition $\mathfrak{u}_{EM}(1)\cap
\mathrm{k}=0$ give the defining representation (with superscript $+$)
and $\newew$-conjugate representation (with superscript $-$) for the doublet
HC matter fields;
\begin{eqnarray}\label{representation}
\bs{T}_{0}^+ &=&\frac{ie}{2\cos{\theta_W}\sin{\theta_W}}\begin{pmatrix}
2\sin^{2}{\theta_W}-1 & 0 \\ 0 & 1 \end{pmatrix},\;\;\;\;\bs{T}_{0}^- =\frac{- ie}{2\cos{\theta_W}\sin{\theta_W}}\begin{pmatrix}
1 & 0 \\ 0 & 2\sin^{2}{\theta_W}-1 \end{pmatrix}\notag \\
\bs{T}^+_{+}&=&\frac{ie}{\sqrt{2}\sin{\theta_W}}\begin{pmatrix}
0 & 1 \\ 0 & 0
\end{pmatrix},\;\;\;\;\bs{T}^-_{+}=\frac{-ie}{\sqrt{2}\sin{\theta_W}}\begin{pmatrix}
0 & 0 \\ 1 & 0 \end{pmatrix}\notag \\
\bs{T}^+_{-}&=&\frac{ie}{\sqrt{2}\sin{\theta_W}}\begin{pmatrix}
0 & 0 \\ 1 & 0 \end{pmatrix},\;\;\;\;\bs{T}^-_{-}=\frac{-ie}{\sqrt{2}\sin{\theta_W}}\begin{pmatrix}
0 & 1 \\ 0 & 0
\end{pmatrix}\notag\\
\bs{Q}^+ &=&ie\begin{pmatrix} 1 & 0 \\ 0 &
0\end{pmatrix},\;\;\;\;\bs{Q}^- =-ie\begin{pmatrix} 0 & 0 \\ 0 &
1\end{pmatrix},
\end{eqnarray}
where $e$ is the electric charge and $\theta_W$ is the Weinberg angle defined by
\begin{eqnarray}
  &g_1^2g_2^2/(g_1^2+g_2^2)&=:e^2\nonumber\\
  &g_1^2&=:\frac{e^2}{\sin^2\theta_W}\nonumber\\
  &(g_1^2+g_2^2)&=\frac{e^2}{\sin^2\theta_W\cos^2\theta_W}\;\;.
\end{eqnarray}
For the $^+$-representation ($^-$-representation) $r(s)$, $t$, and $u(v)$ were absorbed into $g_Q$, $g_W$, and
$g_Z$. This EW representation is identical to the SM
lepton doublet representation --- which is not surprising given the nature of the HC doublet construction.

The $1$-dimensional representations for the charged right-handed iquarks are
obtained by taking the trace of the $2$-dimensional representations
while maintaining proper normalization. For $h_R^{\pm}$ it amounts to $\bs{q}^\pm:=\mathrm{tr}(\bs{Q}^\pm)$ and $\bs{t}_0^\pm:=\mathrm{tr}(\bs{T}^\pm_0)$. Meanwhile, the electrically neutral $\xi_R^0, \chi_R^0$ stem from  the trivial representation and its conjugate.\footnote{Might $\xi_R^0, \chi_R^0$ furnish the determinant representation instead?}

This representation together with (\ref{basis}) implies that the structure coefficients $c_i'$ (in the defining representation) are functions of $(g_1,g_2)$ such that $c'_i(g_1,g_2)\neq0$ for all $i\in\{1,2\}$ for generic values $(g_1,g_2)$.  This has two important consequences for the generators of the Cartan subalgebra: 1) none of the Cartan generators can be proportional to the identity element, and 2) their norms are functions of \emph{both} coupling constants. Notice that $\mathrm{span}_{\C/\sim}\{\bs{T}_0,\bs{T}_+,\bs{T}_-\}$ does not generate $\mathfrak{su}(2)$ except when $g_2\rightarrow0$. Evidently, $\mathfrak{u}_{EW}(2)/\mathfrak{u}_{EM}(1)\not\cong \mathfrak{su}(2)$.\footnote{SM convention has it that isospin and hypercharge are distinguished quantum numbers \emph{before} spontaneous symmetry breaking (SSB) while electric charge $e$ and the mixing angle $\theta_W$ are preferred \emph{after} SSB. Of course electric charge and $\theta_W$ are running couplings. As $g_2/g_1$ varies in the range $0< g_2/g_1< 1$, the character of the neutral bosons' interactions changes. In the limit $g_2\rightarrow0$, electric charge vanishes and $\bs{Q}^\pm$ no longer participates in interactions. Meanwhile, $\bs{T}_0$  and $\bs{T}_\pm$ reduce to the standard fundamental $SU(2)$ representation characterized by isospin. At the other extreme, when $g_2/g_1\rightarrow1$ the relevant quantum number is electric charge since in this case $\theta_W\rightarrow\pi/4$ and $\bs{T}_0^\pm\rightarrow\bs{Q}_\mp$. The point is, electric charge becomes more prominent as energy \emph{increases} while isospin and hypercharge become more prominent as energy \emph{decreases}.}

\begin{remark} Of course our assumed multiplet composition of the HC (with one charged and one neutral component) dictates the representation \emph{(\ref{representation})}. And in general one could construct a different representation consistent with the Cartan decomposition that contains the identity element as one of its generators. But then one of the $c'_i(g_1,g_2)$ would have to vanish identically (for all $(g_1,g_2)$) before SSB and the symmetry group would look like $SU_I(2)\times U_Y(1)$. This, of course, is the starting point of the SM before SSB. Somehow, then, the dynamics of SSB would have to alter the functions $c'_i(g_1,g_2)$ in \emph{(\ref{basis})}. This is a plausible and reasonable picture.

However, an alternative picture is that the dynamics that induce SSB do nothing to change the functional form of the $c'_i(g_1,g_2)$; and, owing to the $U(2)$ invariance, neither of them vanishes identically at energy scales above SSB --- since otherwise this would distinguish a particular direction in the Cartan subalgebra. Equivalently, we insist that $[\Bold{\lambda}_\pm,\Bold{\lambda}_\mp]$ spans the entire Cartan subalgebra. Physically, this means all $U(2)$ bosons interact with each other.

Accordingly, the first equation in \emph{(\ref{basis})} then leads to four conditions
\begin{equation}
[\bs{T}_\pm,\bs{T}_\mp]=\pm c_1'\bs{Q}\pm c_2'\bs{T}_0\Rightarrow
\left\{\begin{array}{c}
c_1'=i\frac{t^2(u+v)}{(ru-sv)}\neq0 \\
c_2'=i\frac{t^2(r+s)}{(ru-sv)}\neq0
\end{array}\right.
\Rightarrow
\left\{\begin{array}{c}
  t\neq0\\
  r+s\neq0\\
  u+v\neq0 \\
  ru-sv\neq0
\end{array}\right.\;.
\end{equation}
Together with $g(\bs{T}_\pm,\bs{Q})=0$ and $g(\bs{T}_0,\bs{Q})=0$, this yields two possible solutions
\begin{equation}
\left(u/v=\frac{g_2^2-g_1^2}{g_1^2+g_2^2}\,,\,\,s=0\right)\;\;\;\;\mathrm{and}\;\;\;\;
\left(r/s=-\frac{g_2^2(u-v)+g_1^2(u+v)}{g_2^2(v-u)+g_1^2(u+v)}\right)\;.
\end{equation}
The first solution on the left is just the representation  \emph{(\ref{representation})} already found. The second solution on the right is under-determined. For the case $u=v$, we have $r/s=-1$ which then implies $\mathrm{span}_{\C/\sim}\{\bs{T}_+,\bs{T}_-,\bs{T}_0,\bs{Q}\}\cong \mathfrak{su}(2)\oplus \mathfrak{u}(1)$ with respect to the inner product and this is unacceptable: besides this violates $r+s\neq0$. The case $u\neq v$ is likewise unacceptable because we don't recover $\mathfrak{su}(2)$ when $g_2\rightarrow0$ unless $u=0$ (or $v=0$). So the second possibility is ruled out unless $u=0$: but then it is the same as the first solution with $(r,s)\leftrightarrow(v,u)$.

Hence, orthogonality together with $\mathrm{span}_{\C/\sim}\{\bs{T}_+,\bs{T}_-,\bs{T}_0,\bs{Q}\}\not\cong \mathfrak{su}(2)\oplus \mathfrak{u}(1)$  and the stipulation\footnote{This ensures the inner product remains finite as $g_2\rightarrow 0$ (which is required for a sensible theory) and that the $g_1$ coupling constants appearing in the SM and our proposed n-SM are equivalent; which means consistent normalizations are maintained between the two theories and they can be compared directly.} that $\mathrm{span}_{\C/\sim}\{\bs{T}_+,\bs{T}_-,\bs{T}_0,\bs{Q}\}|_{g_2\rightarrow0}\cong \mathfrak{su}(2)$ uniquely determine the representation \emph{(\ref{representation})} which in turn dictates the multiplet composition of the HC. We will adopt this picture as a viable, if not more natural, alternative to the conventional view.

For reference, we record the $^\pm$-representation-induced basis for the (evidently reducible) adjoint representation
\begin{eqnarray}
Ad_{\mathbf{\lambda}_+}&=&i e\left(
                   \begin{array}{cccc}
                     0 & 0 & 0 & 0 \\
                     0 & 0 & -\cot\theta_W & 1 \\
                     \cot\theta_W & 0 & 0 & 0 \\
                     -1 & 0 & 0 & 0 \\
                   \end{array}
                 \right)\;\;\;\;
Ad_{\mathbf{\lambda}_-}=i e\left(
                   \begin{array}{cccc}
                     0 & 0 & \cot\theta_W & -1 \\
                     0 & 0 & 0 & 0 \\
                     0 & -\cot\theta_W & 0 & 0 \\
                     0 & 1 & 0 & 0 \\
                   \end{array}
                 \right)\notag\\
Ad_{\mathbf{\lambda}_1}&=&i e\left(
                   \begin{array}{cccc}
                     -\cot\theta_W & 0 & 0 & 0 \\
                     0 & \cot\theta_W & 0 & 0 \\
                     0 & 0 & 0 & 0 \\
                     0 & 0 & 0 & 0 \\
                   \end{array}
                 \right)\;\;\;\;\;\;
Ad_{\mathbf{\lambda}_2}=i e\left(
                   \begin{array}{cccc}
                     1 & 0 & 0 & 0 \\
                     0 & -1 & 0 & 0 \\
                     0 & 0 & 0 & 0 \\
                     0 & 0 & 0 & 0 \\
                   \end{array}
                 \right)
\end{eqnarray}
which produces a non-diagonal and degenerate quadratic Casimir (with respect to \emph{(\ref{inner product})}) represented by
\begin{equation}
 C_{Ad}
 :=\sum_{\sigma,\rho}g_{\sigma\rho}\,\rho_e'(\Bold{\lambda}_\sigma)\cdot\rho_e'(\Bold{\lambda}_\rho)^\dag
 = g_1^2\left(\begin{array}{cccc}
                     1 & 0 & 0 & 0 \\
                     0 & 1 & 0 & 0 \\
                     0 & 0 & \cos^2\theta_W& -\sin\theta_W\cos\theta_W \\
                     0 & 0 &  -\sin\theta_W\cos\theta_W& \sin^2\theta_W \\
                   \end{array}\right)\;.
\end{equation}

\end{remark}

\begin{remark}
What happens if we define different HC that possess the quark electric charges $2/3,-1/3$ of the SM and re-run the analysis? In this case, orthogonality gives
\begin{equation}
\bs{Q}^+_{quark}=ie\left(
                         \begin{array}{cc}
                           2/3 & 0 \\
                           0 & -1/3\\
                         \end{array}
                       \right)\,;\;\;\;\;\;
\bs{T}^+_{0_{quark}}=\frac{i(g_1^2+3g_2^2)}{2\sqrt{g_1^2+9g_2^2}}\left(
                         \begin{array}{cc}
                           -\frac{g_1^2-3g_2^2}{g_1^2+3g_2^2} & 0 \\
                           0 & 1\\
                         \end{array}
                       \right)\;.
\end{equation}
The problem is, because of the numerical factors in front of $g_2^2$, we get for $\theta_W\simeq.23$
\begin{equation}
 \bs{T}^+_{0_{quark}}-\bs{T}_0^+\simeq ie\,\mathrm{diag}(-2.48,4)
\end{equation}
which clearly disagrees with experiment. In fact, the two representations agree only when $\theta_W=0$ which just takes us back to $\mathfrak{su}_{I}(2)\oplus \mathfrak{u}_{Y}(1)$ and the SM before SSB.\footnote{Note, however, that after SSB we no longer have $\bs{T}^+_{0_{quark}}$ orthogonal to $\bs{Q}^+_{quark}$ --- even if we use the Killing inner product instead of (\ref{inner product}).}

Our insistence that $c'_i(g_1,g_2)\neq0$ for all $i\in\{1,2\}$ and generic $(g_1,g_2)$ is a crucial departure from the SM. Without it, the discarded solution found in the previous remark becomes valid; and the EW representation of HC can be constructed exactly as in the SM by specifying isospin and hypercharge quantum numbers --- in which case iquarks no longer posses integer electric charge. So if we allow $c'_i(g_1,g_2)=0$ for some $i\in\{1,2\}$, the only thing $U_{EW}(2)$ brings to the table is the Gell-Mann/Nishijima relation as previously discussed.
\end{remark}

Repeating the exercise for $SU_S(3)$ using the inner product (restricting to $g_1^2> g_2^2$)
\begin{equation}
  g(\bs{\Lambda}_{\alpha},\bs{\Lambda}_{\beta})
  :=-1/6\left[3\tilde{g}_1^{-2}\mathrm{tr}(\bs{\Lambda}_{\alpha}\bs{\Lambda}^\dag_{\beta})
  +(\tilde{g}_2^{-2}-\tilde{g}_1^{-2})\mathrm{tr}\bs{\Lambda}_{\alpha}
  \cdot\mathrm{tr}\bs{\Lambda}^\dag_{\beta}\right]
  =\frac{-1}{2\tilde{g}_1^2}\mathrm{tr}(\bs{\Lambda}_{\alpha}\bs{\Lambda}^\dag_{\beta})\;,
\end{equation}
 \begin{samepage}the $Ad$-invariant inner product on the subalgebra $\mathfrak{su}_S(3):=\mathrm{span}_{\C/\sim}\{\bs{\Lambda}_{\alpha}\}$ is
\begin{equation}
  \left(g_{\alpha\beta}\right)=
  \left(
                 \begin{array}{cccccccc}
                   {g_{s1}}^{-2} & 0 & 0 & 0 & 0 & 0 & 0 & 0 \\
                   0 & {g_{s1}}^{-2} & 0 & 0 & 0 & 0 & 0 & 0 \\
                   0 & 0 & {g_{s2}}^{-2} & 0 & 0 & 0 & 0 & 0 \\
                   0 & 0 & 0 & {g_{s2}}^{-2} & 0 & 0 & 0 & 0 \\
                   0 & 0 & 0 & 0 & {g_{s3}}^{-2} & 0 & 0 & 0 \\
                   0 & 0 & 0 & 0 & 0 & {g_{s3}}^{-2} & 0 & 0 \\
                   0 & 0 & 0 & 0 & 0 & 0 & {g_{h1}}^{-2} & 0 \\
                   0 & 0 & 0 & 0 & 0 & 0 & 0 & {g_{h2}}^{-2} \\
                 \end{array}
               \right)
\end{equation}
with ${g_{si}}^{-2}:=g({(\bs{\Lambda}_i)}_\pm,{(\bs{\Lambda}_i)}_\pm)$, ${g_{h1}}^{-2}:=g(\bs{\Lambda}_{1},\bs{\Lambda}_{1})$, and ${g_{h2}}^{-2}:=g(\bs{\Lambda}_{2},\bs{\Lambda}_{2})$.
\end{samepage}

An explicit defining representation is given by
\begin{eqnarray}
&& \bs{S}^+_1:=\rho'({(\bs{\Lambda}_1)}_+)= ig_{s1}\begin{pmatrix}
0 & 1 & 0\\0 & 0 & 0\\0&0&0
\end{pmatrix}\;\;\;\;\bs{S}^-_1:=\rho'({(\bs{\Lambda}_1)}_-)= ig_{s1}\begin{pmatrix}
0&0&0\\1&0&0\\0&0&0
\end{pmatrix}
\notag\\
&&\bs{S}^+_2:=\rho'({(\bs{\Lambda})}_+)= ig_{s2}\begin{pmatrix}
0&0&1\\0&0&0\\0&0&0
\end{pmatrix}\;\;\;\;\;\bs{S}^-_1:=\rho'({(\bs{\Lambda}_1)}_-)=i g_{s2}\begin{pmatrix}
0&0&0\\0&0&0\\1&0&0
\end{pmatrix}
\notag\\
&&\bs{S}^+_3:=\rho'({(\bs{\Lambda}_3)}_+)= ig_{s3}\begin{pmatrix}
0&0&0\\0&0&1\\0&0&0
\end{pmatrix}\;\;\;\;\bs{S}^-_3:=\rho'({(\bs{\Lambda}_3)}_-)= ig_{s3}\begin{pmatrix}
0&0&0\\0&0&0\\0&1&0
\end{pmatrix}
\notag\\
&&\bs{H}_1:=\rho'({\bs{\Lambda}_7})= ig_{h1}\begin{pmatrix}
-1&0&0\\0&1&0\\0&0&0
\end{pmatrix}\;\;\;\;\;\;\bs{H}_2:=\rho'({\bs{\Lambda}_8)}=i g_{h2}
\begin{pmatrix}
1&0&0\\0&1&0\\0&0&-2
\end{pmatrix}
\end{eqnarray}
 where $g_{si}=g_{h1}=g_{h2}=\tilde{g}_1$ is the strong coupling constant.

 With this choice, the inner product in the defining representation becomes
\begin{equation}
  \left(g_{\alpha\beta}\right)=
  1/2\left(
                 \begin{array}{cccccccc}
                   1 & 0 & 0 & 0 & 0 & 0 & 0 & 0 \\
                   0 & 1 & 0 & 0 & 0 & 0 & 0 & 0 \\
                   0 & 0 & 1 & 0 & 0 & 0 & 0 & 0 \\
                   0 & 0 & 0 & 1 & 0 & 0 & 0 & 0 \\
                   0 & 0 & 0 & 0 & 1 & 0 & 0 & 0 \\
                   0 & 0 & 0 & 0 & 0 & 1 & 0 & 0 \\
                   0 & 0 & 0 & 0 & 0 & 0 & 2 & 0 \\
                   0 & 0 & 0 & 0 & 0 & 0 & 0 & 6 \\
                 \end{array}
               \right)\;.
\end{equation}
This, of course, is not the standard representation or normalization. But it is convenient because the generators are all imaginary and it yields integer intrinsic charges.\footnote{As with the previous case of $\mathfrak{u}_{EW}(2)$, to get Hermitian matrices we take $\mathfrak{su}_S(3)\equiv\mathrm{span}_{\C/\sim}\{\bs{S}_i^\pm,\bs{H}_i\}$. Standard normalization is readily achieved by putting $\sqrt{2}g_{h1}=\sqrt{6}g_{h2}=\tilde{g}_1$.} Explicitly, gluons posses  the strong charges (as factors of $\tilde{g}_1$)  ${(\bs{\Lambda}_1)_\pm}\rightarrow{(\mp2,0)}$, ${(\bs{\Lambda}_2)_\pm}\rightarrow{(\mp1,\mp3)}$, ${(\bs{\Lambda}_3)_\pm}\rightarrow{(\pm1,\mp3)}$, $\bs{\Lambda}_7\rightarrow{(0,0)}$, and ${\bs{\Lambda}_8}\rightarrow{(0,0)}$. Similarly, the three matter eigenfields in the defining representation posses strong charges $(\mp1,\mp1)$, $(\pm1,\mp1)$, and $(0,\pm2)$.

\subsection{The Lagrangian density}
The Yang-Mills, lepton, and Higgs contributions to the n-SM Lagrangian
density are identical (up to our non-standard  inner product and normalization) to the SM so we won't display them. But, according to our previous discussion, the inequivalent iquark irreps of the direct product group
include the combinations $(\mathbf3,\mathbf2)$,
$(\mathbf3,{\mathbf2}^{\bs{c}})$, $(\mathbf3,\mathbf1^+)$,
$(\mathbf3,\mathbf1^-)$, and $(\mathbf3,\mathbf1^0)$ along with
 corresponding anti-particle combinations. Their contribution to the Lagrangian density is
\begin{eqnarray} \label{lagrangian}
\mathcal{L}_{\text{iquark}}&=&i\sum_{s}
 {\kappa^+(\overline{\Bold{H}_{{\mathrm{L}},s}^+}\D^+
 \Bold{H}^+_{{\mathrm{L}},s}+\overline{\Bold{h}_{{\mathrm{R}},s}^+}\D^+
 \Bold{h}^+_{{\mathrm{R}},s}+\overline{\Bold{\xi}_{\mathrm{R,s}}^0}\D\Bold{\xi}^0_{\mathrm{R},s})}\nonumber\\
 &&\hspace{.4in}
 +\kappa^-(\overline{\Bold{H}_{{\mathrm{L}},s}^-}\D^-
 \Bold{H}^-_{{\mathrm{L}},s}+\overline{\Bold{h}_{{\mathrm{R}},s}^-}\D^-
 \Bold{h}^-_{{\mathrm{R}},s}+\overline{\Bold{\chi}_{\mathrm{R,s}}^0}\D\Bold{\chi}^0_{\mathrm{R},s})
\\ \nonumber\\
\mathcal{L}_{\text{Yukawa}}&=&-\sum_{s,t} \kappa^+(m^+_{st}
\overline{\Bold{h}_{{\mathrm{R}},s}^+}{\mathit{\Phi}^+}^\dag{\Bold{H}}^+_{{\mathrm{L}},t}
+ n^+_{st}
\overline{\Bold{H}_{{\mathrm{L}},s}^+}
\mathit{\Phi}^+\Bold{\xi}^0_{{\mathrm{R}},t})\nonumber\\
&&\hspace{.4in}
+\kappa^-(m^-_{st}
\overline{\Bold{h}_{{\mathrm{R}},s}^-}{\mathit{\Phi}^-}^\dag{\Bold{H}}^-_{{\mathrm{L}},t} + n^-_{st}
\overline{\Bold{H}_{{\mathrm{L}},s}^-}
\mathit{\Phi}^-\Bold{\chi}^0_{{\mathrm{R}},t})+h.c.
\end{eqnarray}
where $s,t$ label iquark generation and $h.c.$ means Hermitian conjugate.

The covariant derivatives are
\begin{eqnarray} \label{covariant derivative}
\D^+ \Bold{H}^{+}_{\mathrm{L}} &=& \left(\not\!{\partial}+
\not\!\!{W}^{+}\bs{T}^{+}_{+}+ \not\!\!{W}^{-}\bs{T}^{+}_{-}+
\not\!\!{Z}^{0}\bs{T}^{+}_{0}
+ \not\!\!{A}\bs{Q}^{+}+\not\!G\bs{\Lambda}\right)\Bold{H}^{+}_{\mathrm{L}}\notag \\\nonumber\\
\D^- \Bold{H}^{-}_{\mathrm{L}} &=& \left(\not\!{\partial}+
\not\!\!{{W}^{+}}^*\bs{T}^{-}_{-}+
\not\!\!{{W}^{-}}^*\bs{T}^{-}_{+}+ \not\!\!{Z}^{0}\bs{T}^{-}_{0}
+ \not\!\!{A}\bs{Q}^{-}+\not\!G\bs{\Lambda}\right)\Bold{H}^-_{\mathrm{L}}\nonumber\\
\nonumber\\
 \D^{\pm} \Bold{h}^{\pm}_{\mathrm{R}}
&:=&
\mathrm{tr}\mathit[\D^{\pm}]\Bold{h}^{\pm}_{\mathrm{R}}\,,\;\;\;\;
\D\Bold{\xi}^0_{\mathrm{R}} =
\bigl(\not\!\partial+\not\!{G}\bs{\Lambda}\bigr)\Bold{\xi}^0_{\mathrm{R}}\,,\;\;\;\;
\D\Bold{\chi}^0_{\mathrm{R}} =
\bigl(\not\!\partial+\not\!{G}\bs{\Lambda}\bigr)\Bold{\chi}^0_{\mathrm{R}}\;,
\end{eqnarray}
where the trace is only over $\newew$ indices.  The matrices $m^\pm_{st}$ and
$n^\pm_{st}$ are generation-mixing complex matrices, and
$\mathit{\Phi}^+$ is the Higgs field
\begin{equation}
  \mathit{\Phi}^+:=\begin{pmatrix}
  \phi^+ \\ \phi^0
  \end{pmatrix}\;.
\end{equation}

Clearly $\kappa^\pm$ in the Yukawa term can be absorbed into the mixing matrices so they play no role in the Yukawa term which can then be expressed in the unitary gauge as
\begin{eqnarray}\label{Yukawa}
\mathcal{L}_{\text{Yukawa}}
&=&-\sum_{s,t}
(m_{st}^+\overline{\Bold{h}_{{\mathrm{R}},s}^+}{\mathit{\Phi}^+}^\dag{\Bold{H}}^+_{{\mathrm{L}},t}
+m_{st}^-\overline{\Bold{h}_{{\mathrm{R}},s}^-}{\mathit{\Phi}^-}^\dag{\Bold{H}}^-_{{\mathrm{L}},t})\nonumber\\
&&\hspace{.4in}+
(n_{st}^+\overline{\Bold{H}_{{\mathrm{L}},s}^+}
\mathit{\Phi}^+\Bold{\xi}^0_{{\mathrm{R}},t}+n_{st}^-\overline{\Bold{H}_{{\mathrm{L}},s}^-}
\mathit{\Phi}^-\Bold{\chi}^0_{{\mathrm{R}},t})+h.c.\nonumber\\
&=&-\sum_{s,t}
(m_{st}^+\overline{\Bold{h}_{{\mathrm{R}},s}^+}{\Bold{h}}^+_{{\mathrm{L}},t}
+n_{st}^+\overline{\Bold{\xi}_{{\mathrm{L}},s}^0}
\Bold{\xi}^0_{{\mathrm{R}},t})\phi^0\nonumber\\
&&\hspace{.4in}+
(m_{st}^-\overline{\Bold{h}_{{\mathrm{R}},s}^-}{\Bold{h}}^-_{{\mathrm{L}},t}+n_{st}^-\overline{\Bold{\chi}_{{\mathrm{L}},s}^0}
\Bold{\chi}^0_{{\mathrm{R}},t})\phi^0+h.c.\;\;.
\end{eqnarray}
 Hence, unitary transformations in generation space yield real, diagonal mass matrices for $m_{st}^\pm$ and $n_{st}^\pm$; and the HC can be redefined in terms of the mass eigenfields according to
\begin{equation}\label{new HC}
\Bold{H}^+:=\left(
         \begin{array}{c}
           V^+_{(m)}h^+ \\
           V^+_{(n)}\xi^0 \\
         \end{array}
       \right)^A\Bold{e}_A\;,\;\;\;\;\;\;\Bold{H}^-:=\left(
         \begin{array}{c}
           V^-_{(n)}\chi^0 \\
           V^-_{(m)}h^-\\
         \end{array}
       \right)^A\Bold{e}_A
\end{equation}
where $V_{{(m)}}^\pm$ and $V_{{(n)}}^\pm$ are unitary CKM-like generation-mixing matrices and we have abused notation by re-using $h^+,h^-,\xi^0,\chi^0$ to denote mass eigenfields.

A few remarks are in order:
\begin{itemize}
\item In the lepton sector, putting $\Bold{H}^{-}\rightarrow \tau_1{\Bold{H}^{+}}^\ast$  transforms $\mathcal{L}_{\text{lepton}}$ in the form of (\ref{lagrangian}) into the usual SM quark Lagrangian density. This happens because
    \begin{equation}
    i\overline{\Bold{H}^{-}}\slashed{D}^-_{lepton}\Bold{H}^{-}
    =(i\overline{\Bold{H}^{-}}\slashed{D}^-_{lepton}\Bold{H}^{-})^\ast
    \rightarrow i\overline{\Bold{H}^{+}}\slashed{D}^+_{lepton}\Bold{H}^{+}+\mathrm{total\,derivative}
    \end{equation}
    and we get $\mathcal{L}_{\text{lepton}}
    =(\kappa^++\kappa^-)i\overline{\Bold{H}^{+}}\slashed{D}^+_{lepton}\Bold{H}^{+}$. In contrast, in the iquark sector $\slashed{D}^\pm$ contains $SU_S(3)$ terms $\slashed{G}\bs{\Lambda}$ that do not transform under $U_{EW}(2)$-conjugation so $(\tau_1){\slashed{D}^-}(\tau_1)^{-1}\neq{\slashed{D}^+}^\ast$ and then $(i\overline{\Bold{H}^{-}}\slashed{D}^-\Bold{H}^{-})^\ast
    \neq i\overline{\Bold{H}^{+}}\slashed{D}^+\Bold{H}^{+}+\mathrm{total\,derivative}$. The two theories, with just $\Bold{H}^{+}$ v.s. $\Bold{H}^{+}$ and $\Bold{H}^{-}$, are not equivalent in the iquark sector.

\item  Re-scaling the iquark fields cannot
cancel the relative scale difference between $\Bold{H}^{+}$ and
$\Bold{H}^{-}$ since they are $\newew$-conjugate to each other
(unless $\kappa^+=\kappa^-$). Consequently, these factors are not
trivial and it was already shown that their ratio is not altered by renormalization. The
effect of the constants $\kappa^+$ and $\kappa^-$ is to re-scale
the charge $e$ in (\ref{representation}). Note that the $SU_S(3)$ coupling strengths are not altered as long as
$\kappa^++\kappa^-=1$. That is, $\kappa^++\kappa^-=1$ guarantees $SU_S(3)$ intrinsic and extrinsic charge equality under $U(2)$ gauge transformations. Consequently, strong-force interactions are not disturbed by weak-force coupling renormalization.

\item
$\mathcal{L}_{\mathrm{iquark}}$ is not invariant under distinct $U_{EW}(2)$ gauge transformations of $\Bold{H}^{+}$ and $\Bold{H}^{-}$ for two reasons. First, the Yukawa term forbids it. Second, the $\newew$-conjugate representation is reached through the adjoint action of $\tau_1$ with $\det{\tau_1}=-1$ and $\tau_1^2=1$, and this signals a large gauge transformation.\footnote{Consider an infinitesimal gauge transformation $U(x)=
    \bs{1}+i\alpha^\pm(x)\Bold{\lambda}_\pm+i\alpha^i(x)\Bold{\lambda}_i+O(\alpha^2)$. For the defining representation, make use of the identity $\det A=1/2\left((\mathrm{tr}\, A)^2-\mathrm{tr}\, A^2\right)$ valid  in  two dimensions. Then, to first order in $\alpha$ calculate $(\mathrm{tr}\, U(x))^2=(\mathrm{tr}\,\bs{1}+i\alpha^i(x)\mathrm{tr}\,\Bold{\lambda}_i)^2\simeq
    (\mathrm{tr}\,\bs{1})^2+2i\alpha^i(x)\mathrm{tr}\,\Bold{\lambda}_i$ as well as $\mathrm{tr}\,(U(x)^2)\simeq\mathrm{tr}\,\left(\bs{1}
    +2i\alpha^\pm(x)\Bold{\lambda}_\pm+2i\alpha^i(x)\Bold{\lambda}_i\right)=\mathrm{tr}\,(\bs{1}
    +2i\alpha^i(x)\Bold{\lambda}_i)$. Hence, $\det U(x)\simeq (\mathrm{tr}\,\bs{1})^2-\mathrm{tr}\,\bs{1}>0$ to first order in $\alpha$.  Conclude that a gauge transformation homotopic to the identity implies $\det U(x)>0$.} Consequently, any gauge transformation $R$ on $\mathrm{span}_\R\{\Bold{e}^a\}$ induces a conjugate transformation on $\mathrm{span}_\R\{\Bold{e}^a_{\bs{c}}\}$ given by ${R}^{\bs{c}}=(\tau_1) R^\ast(\tau_1)^{-1}$. But $\tau_1,R^\ast\in U(2)$ and the gauge algebra is closed with respect to the adjoint action, so ${R}^{\bs{c}}$ is also a gauge transformation --- different but not independent.

\item There is an approximate discrete symmetry under $\Bold{H}^{+}\leftrightarrow\Bold{H}^{-}$ as long as $\kappa^+/\kappa^-$ is non-trivial and not too small. In consequence, there is an approximate global symmetry $U(2)\times\mathbb{Z}_2\simeq SU(2)\times U(1)$, and in the limit of vanishing masses this extends to $SU(2)_L\times SU(2)_R\times U(1)_L\times U(1)_R$ (just like in the SM).

\item The $\xi_R^0,\,\chi_R^0$ fields completely decouple from the $\newew$
gauge bosons. However, they do couple to the $SU_S(3)$ gauge
bosons. They also have an induced mass due to the Higgs
interaction included in $\mathcal{L}_{\text{Yukawa}}$ (if $n_{st}\neq0$).

\item The Yukawa and EW terms in the n-SM include twice as many mass parameters ($m_{st}^\pm$ and $n_{st}^\pm$) and CKM parameters ($V^+:={V_{(m)}^+}^\dag V_{(n)}^+$ and $V^-:={V_{(m)}^-}^\dag V_{(n)}^-$) due to the elementary content being $(\mathbf{3},\mathbf{2})\oplus(\mathbf{3},{\mathbf{2^c}})$. Symmetry under electric charge conjugation implies ${V^+}^\dag=-V^-$, but since individual quark/iquark currents are not observed the extra mass parameters don't seem to be inferable without additional assumptions. One possibility rests on the model proposed in \S 4 which implies a relation between iquarks and leptons that would (in principle) relate the iquark/lepton masses; thus reducing the free mass parameters to the same as the SM (since lepton/anti-lepton pairs possess identical masses).

\end{itemize}

\subsection{Comparing to the SM}\label{SM comparison}
Recall that $e^a\cdot e^a=e^a_{\bs{c}}\cdot e^a_{\bs{c}}$ together with (\ref{currents 2}) imply each eigenfield contribution to the $SU_S(3)$ current  gets multiplied by the harmless constant $\kappa^++\kappa^-=1$. In other words, the strong interaction cannot distinguish the difference between $\pm$ iquarks. In consequence, although there are $4\times3$ elementary iquarks, the strong interaction in hadrons ``effectively sees" only $2\times3$ iquarks carrying the total requisite (same as the SM) strong charges. Hence, the QCD sector of the n-SM agrees precisely with the SM.

It suffices to compare the quark versus iquark content of the Yukawa and EW sectors since everything else is unaltered. The Yukawa term describes the same Higgs interaction as the SM; albeit with different elementary fields and double the mass parameters. To compare EW sectors, start with the n-SM EW currents.

\subsubsection{Currents and Anomalies}
Using (\ref{matter fields}), (\ref{representation}),
(\ref{lagrangian}), and (\ref{new HC}) the $\newew$ currents for each iquark
generation are
\begin{eqnarray} \label{currents}
j^{0(Z)}_{\mu}&=&\frac{e}{2\sin{\theta_W}\cos{\theta_W}}
 \biggl[\kappa^+\Bigl(2\sin^{2}{\theta_W}-1\Bigr)
\overline{h^{+}_{\mathrm{L}}}\gamma_{\mu}h^{+}_{\mathrm{L}}
\notag\\
 & & -\kappa^-\Bigl(2\sin^{2}{\theta_W}-1\Bigr)
\overline{h^{-}_{\mathrm{L}}}\gamma_{\mu}h^{-}_{\mathrm{L}}\notag\\
 & & +\kappa^+2\sin^{2}{\theta_W}\overline{h^{+}_{\mathrm{R}}}\gamma_{\mu}
h^{+}_{\mathrm{R}}+\kappa^+\overline{\xi^{0}_{\mathrm{L}}}
\gamma_{\mu}\xi^{0}_{\mathrm{L}}\notag\\
 & & -\kappa^-\overline{\chi^{0}_{\mathrm{L}}}
\gamma_{\mu}\chi^{0}_{\mathrm{L}}
-\kappa^-2\sin^{2}{\theta_W}\overline{h^{-}_{\mathrm{R}}}
\gamma_{\mu}h^{-}_{\mathrm{R}}\biggr], \notag \\\notag \\
j^{0(A)}_{\mu}&=&
\kappa^+e\overline{h^{+}}\gamma_{\mu}h^{+}-\kappa^-e \overline{h^{-}}
\gamma_{\mu}h^{-},\notag\\\notag \\
j^{-}_{\mu}&=&\frac{e}{\sqrt{2}\sin{\theta_W}} \biggl[
\kappa^+\overline{h^{+}_{\mathrm{L}}}\gamma_{\mu}V^+\xi^{0}_{\mathrm{L}}
-\kappa^-\overline{\chi^{0}_{\mathrm{L}}}{V^-}^\dag\gamma_{\mu}
h^{-}_{\mathrm{L}}\biggr],\notag \\\notag \\
j^{+}_{\mu}&=&\frac{e}{\sqrt{2}\sin{\theta_W}} \biggl[
\kappa^+\overline{\xi^{0}_{\mathrm{L}}}{V^+}^\dag\gamma_{\mu}h^{+}_{\mathrm{L}}
-\kappa^-\overline{h^{-}_{\mathrm{L}}}\gamma_{\mu}
{V^-}\chi^{0}_{\mathrm{L}}\biggr]
\end{eqnarray}
where we used ${W^\pm}^*=W^\mp$ for the $\kappa^-$ terms in the
two charged currents, $V^+:={V_{(m)}^+}^\dag V_{(n)}^+$, $V^-:={V_{(m)}^-}^\dag V_{(n)}^-$, and summation over $SU_S(3)$ indices is implicit.

As is well known, for a consistent quantum version of this model
to exist, the anomalies associated with these currents must cancel
the lepton anomalies. Because the iquarks furnish the same EW
representation as the leptons and because there are three degrees of freedom of each, one would not expect the anomalies in this model
to cancel trivially.

To check this, it must be kept in mind that the $U_{EM}(1)$ quantities which enter
into the anomaly calculation are not the intrinsic electric
charges of the matter fields, per se, but the coupling strengths
in the photon-matter field current $j^{0(A)}_\mu$. The
$U_{EM}(1)$, $U_{EW}(2)$, and $SU_S(3)$ contributions  of the
left-handed matter fields are given in Table \ref{table}.

\begin{table}[h]
\centering
\begin{tabular}{|l| c c c c c c c c c|}
\hline fermions &
\begin{math}(h^{+},\xi^{0})_{\mathrm{L}}\end{math} &
\begin{math}(\chi^{0},h^{-})_{\mathrm{L}}\end{math} &
\begin{math}\overline{h^{+}_\mathrm{R}}\end{math} &
\begin{math}\overline{h^{-}_\mathrm{R}}\end{math} &
\begin{math}\overline{\xi^{0}_\mathrm{R}}\end{math} &
\begin{math}\overline{\chi^{0}_\mathrm{R}}\end{math} &
\;\;\;\;\begin{math}(\nu^{0},l^{-})_{\mathrm{L}}\end{math}&
\begin{math}\overline{l^{-}_\mathrm{R}}\end{math} &
\begin{math}\overline{\nu^{0}_{\mathrm{R}}}(?)\end{math}\\ \hline\hline
\begin{math}U(1)\end{math} & \begin{math}(\kappa^+,0)\end{math} &
\begin{math}(0,-\kappa^-)\end{math} &  \begin{math}-\kappa^+\end{math} &
\begin{math}\kappa^-\end{math} & \begin{math}0\end{math} & \begin{math}0\end{math} &
\;\;\;\;\begin{math}(0,-1)
\end{math} & 1 & 0\\
\begin{math}U(2)\end{math} & 2 & \begin{math}2^{\bs{c}}\end{math} & \begin{math}\overline{1}\end{math} & \begin{math}\overline{1^{\bs{c}}}\end{math} &   \begin{math}\overline{1}\end{math} & \begin{math}\overline{1^{\bs{c}}}\end{math} & \;\;\;\;2 & 1 & 1 \\
\begin{math}SU(3)\end{math} & 3 & 3 & \begin{math}\overline{3}
\end{math} & \begin{math}\overline{3}
\end{math} &\begin{math}\overline{3}
\end{math} & \begin{math}\overline{3}\end{math} & \;\;\;\;1 & 1 & 1 \\
\hline
\end{tabular}
\caption{Anomaly contributions for left-handed fermionic matter
fields.} \label{table}
\end{table}

There are only four cases to check including the gravitational
anomaly \cite{W3}: $[U(2)]^2U(1)$, $[SU(3)]^2U(1)$, $[U(1)]^3$,
and $[G]^2U(1)$. (The $[SU(3)]^3$ case vanishes trivially since the representation is real.) In that order, the relevant terms are
\begin{subequations}
\begin{align}
\sum_{\text{doublets}} p&=3(\kappa^+)
+3(-\kappa^-)+(-1)=0\;,\label{non-trivial}\\
\sum_{\text{triplets}} p&=(\kappa^+) +(-\kappa^-)+(-\kappa^+)+
(\kappa^-)+0+0=0\;,\\
\sum_{\text{all}} p^{3}&=3(\kappa^+)^{3}
+3(-\kappa^-)^{3} +3(-\kappa^+)^{3} +3(\kappa^-)^{3} +(-1)^{3}+(1)^{3}=0\;,\\
\sum_{\text{all}} p&=3(\kappa^+) +3(-\kappa^-)+3(-\kappa^+)
+3(\kappa^-) +(-1)+(1)=0\;,
\end{align}
\end{subequations}
where $ep$ denotes the $U_{EM}(1)$ coupling parameter for the
iquark currents. With the exception of (\ref{non-trivial}), the
anomaly conditions are null rather trivially. From
(\ref{non-trivial}) and the condition $\kappa^++\kappa^-=1$, there
will be no anomaly associated with the gauge symmetries for the
choice
\begin{equation}\label{fractions}
\kappa^+=\frac{2}{3}\;,  \ \ \ \ \  \kappa^-=\frac{1}{3}\;.
\end{equation}

Now turn to the chiral anomaly and the decay rate of $\pi^{0}\rightarrow 2\gamma$ associated with the global chiral transformation
\begin{eqnarray}
\delta_\lambda H^A_+=\lambda\mathbf{\Xi}H_+^A &\ \ \ \ &
\delta_\lambda H^A_-=\lambda\mathbf{\Xi}'H_-^A\;
\end{eqnarray}
where $\mathbf{\Xi}=\left(\begin{array}{cc}
                                i\gamma_5 & 0 \\
                                0 & -i\gamma_5 \\
                              \end{array}
                            \right)$
and $\mathbf{\Xi}'=(\tau_1)\mathbf{\Xi}^\ast(\tau_1)^\dag=\mathbf{\Xi}$. The anomaly is proportional to
\begin{equation}
\mathrm{tr}_{U(2)}\biggl[(\kappa^+ \bs{Q}^{+})^2\mathbf{\Xi}+(\kappa^- \bs{Q}^{-})^2\mathbf{\Xi}'\biggr]\propto \mathrm{tr}_{U(2)}\biggl[(\kappa^+ \bs{Q}^{+}+\kappa^-
\bs{Q}^{-})^{2}\left(
                              \begin{array}{cc}
                                1 & 0 \\
                                0 & -1 \\
                              \end{array}
                            \right)
\biggr]
\end{equation}
which, for three colors and using (\ref{fractions}), yields the SM result
\begin{equation}
3\Bigl(\frac{2}{3}\Bigr)^{2}-3\Bigl(\frac{1}{3} \Bigr)^{2}=1\;.
\end{equation}

\subsubsection{Quarks v.s. iquark}
To make contact with SM phenomenology, we need to associate the conventional
generation-mixed quark mass eigenstates with
$h^{\pm},\,\xi^0,\,\chi^0$. Inspection of (\ref{currents}) suggests that
the familiar fractionally charged quark mass eigenstates should be associated with
a pair of HC. Thus, make the following
correspondence:
\begin{eqnarray}\label{quarks}
\left(
  \begin{array}{c}
    u^{+\frac{2}{3}} \\
   d^{-\frac{1}{3}} \\
  \end{array}
\right)
\leftrightarrow
\left(
      \begin{array}{c}
        h^{+} \\
        \xi^{0} \\
      \end{array}
    \right)
+
\left(
      \begin{array}{c}
        \chi^{0} \\
        h^{-} \\
      \end{array}
    \right)
\end{eqnarray}
where $u$ and $d$ represent up and down quark fields respectively.
More accurately, the neutral quark bilinears are identified with a pair of neutral HC
bilinears
\begin{eqnarray}\label{correspondance 1}
 \overline{u^{+\frac{2}{3}}}\gamma_{\mu}u^{+\frac{2}{3}}&\leftrightarrow
&\left(\overline{h^{+}}\gamma_{\mu}h^{+}
 \,;\,\overline{\chi^{0}}\gamma_{\mu}\chi^{0}\right)
\nonumber\\
 \overline{d^{-\frac{1}{3}}}\gamma_{\mu}d^{-\frac{1}{3}}
 &\leftrightarrow &\left(\overline{\xi^{0}}\gamma_{\mu}\xi^{0}\,;\,
\overline{h^{-}}\gamma_{\mu}h^{-}\right)\;,
\end{eqnarray}
and the charged quark bilinears are identified with charged HC pairs
\begin{eqnarray}\label{correspondance 2}
\overline{u^{+\frac{2}{3}}}\gamma_{\mu}V d^{-\frac{1}{3}} &\leftrightarrow
& \left(\overline{h^{+}}\gamma_{\mu}V^+\xi^{0}
\,;\,\overline{\chi^{0}}{V^-}^\dag\gamma_{\mu}h^{-}\right)
\nonumber\\
\overline{d^{-\frac{1}{3}}}V^\dag\gamma_{\mu}u^{+\frac{2}{3}} &\leftrightarrow
& \left(\overline{\xi^{0}}{V^+}^\dag\gamma_{\mu}h^{+}
\,;\,\overline{h^{-}}\gamma_{\mu}{V^-}\chi^{0}\right)
\end{eqnarray}
where $V$ is the CKM matrix and ${V^+}^\dag=-V^-$.

So, the neutral and charged EW currents can be compared in the two different pictures with the help of (\ref{currents});
\begin{eqnarray}
\tfrac{2}{3}\overline{u^{+\frac{2}{3}}}\gamma_{\mu}u^{+\frac{2}{3}}&\sim&
\tfrac{2}{3}\overline{h^{+}}\gamma_{\mu}h^{+}\;,
\nonumber\\
-\tfrac{1}{3} \overline{d^{-\frac{1}{3}}}
\gamma_{\mu}d^{-\frac{1}{3}}&\sim &-\tfrac{1}{3} \overline{h^{-}}\gamma_{\mu}h^{-};\,
\end{eqnarray}
\begin{eqnarray}
\Bigl(\tfrac{4}{3}\sin^{2}{\theta}-1\Bigr)
\overline{u^{+\frac{2}{3}}_{L}}\gamma_{\mu}u^{+\frac{2}{3}}_{L} &\sim &
\Bigl(\tfrac{4}{3}\sin^{2}{\theta}-\tfrac{2}{3}\Bigr)
\overline{h^{+}_{L}}\gamma_{\mu}h^{+}_{L} -
\tfrac{1}{3}\overline{\chi^{0}_{L}}\gamma_{\mu}\chi^{0}_{L}\;,
\nonumber\\
\Bigl(\tfrac{4}{3}\sin^{2}{\theta}\Bigr)
\overline{u^{+\frac{2}{3}}_{R}}\gamma_{\mu}u^{+\frac{2}{3}}_{R} &\sim &
\Bigl(\tfrac{4}{3}\sin^{2}{\theta}\Bigr)
\overline{h^{+}_{R}}\gamma_{\mu}h^{+}_{R}\;,
\end{eqnarray}
and
\begin{eqnarray}
\overline{u^{+\frac{2}{3}}_{L}}\gamma_{\mu}Vd^{-\frac{1}{3}}_{L} \sim
 \tfrac{2}{3}\overline{h^{+}_{L}}\gamma_{\mu}V^+\xi^0_{L}
+\tfrac{1}{3}\overline{\chi^{0}_{L}}V^+\gamma_{\mu}h^-_{L}
\end{eqnarray}
where we used $V^+=-{V^-}^\dag$. There are analogous relations for
$\overline{d^{-\frac{1}{3}}_{L}}\gamma_{\mu}d^{-\frac{1}{3}}_{L}$ and
$\overline{d^{-\frac{1}{3}}_{L}}V^\dag\gamma_{\mu}u^{+\frac{2}{3}}_{L}$.

To the extent that (\ref{correspondance 1}) and (\ref{correspondance 2}) are justified, the weak
currents in (\ref{currents}) agree precisely with the SM currents.
Graphically, the correspondence associates one-particle
quark currents and their vertex factors with an equivalent two-particle
iquark current whose vertex factor is the sum of the individual vertex
factors of the constituent one-particle currents. Physically, the
correspondence constitutes an average description in the sense
that individual quarks/iquarks cannot be resolved.

Observe, however, that the iquark parton distribution and structure functions of charged lepton--hadron  deep inelastic scattering(DIS) interactions will differ from the standard parton model since a (presumably small) portion of the spin-$1/2$ mass-energy of the hadron comprised of $\xi^0,\,\chi^0$ does not directly interact with the lepton via $U_{EM}(1)$: To the extent that the strong coupling between HC decreases, the spin of the neutral iquarks will not participate in charged lepton DIS. This may have implications regarding the proton spin puzzle (see e.g. \cite[\S5]{BASS}) and EMC effect.

\subsubsection{Hadrons}
With the n-SM EW currents and quark/iquark correspondences in hand, we can compare the SM and n-SM phenomenology, but first we need to exhibit the iquark content of hadrons.

According to (\ref{currents}), iquark EW currents include iquarks belonging to both $\Bold{H}^+$ and $\Bold{H}^-$. This gives a hint about how to assign iquark content to hadrons. Based on the relationship between quark v.s. iquark EW currents and the circumstance that charged weak interactions interchange the up and down components of $\Bold{H}^\pm$, it is convenient to define valence HC composed of up-type and down-type iquark \textit{pairs} in order to characterize hadrons
\begin{eqnarray}\label{valence iquarks}
H^\uparrow_s&:=&h_s^++\chi_s^0=\mathrm{P}^\uparrow(\Bold{H}_s^++\Bold{H}_s^-)\notag\\
H^\downarrow_s&:=&\xi_s^0+h_s^-=\mathrm{P}^\downarrow(\Bold{H}_s^++\Bold{H}_s^-)\;.
\end{eqnarray}
where $\mathrm{P}^\updownarrow$ projects onto the up/down $U_{EW}(2)$ component.

The n-SM currents seem to be telling us that we should think of $H^\updownarrow_s$ like up/down quarks. But will this lead to acceptable spin content when we try to form hadrons? Let's enumerate the spin possibilities: denote the spin content  by $|H^\updownarrow_s\rangle=(|\frac{1}{2},\pm\frac{1}{2}\rangle;|\frac{1}{2},\pm\frac{1}{2}\rangle)$. For meson composites $H^\updownarrow_s\overline{H^\updownarrow_t}
:=\mathrm{P}^\updownarrow\Bold{H}_s^+\overline{\mathrm{P}^\updownarrow\Bold{H}_t^+}
+\mathrm{P}^\updownarrow\Bold{H}_s^-\overline{\mathrm{P}^\updownarrow\Bold{H}_t^-}$, the possible spin combinations are
\begin{equation}
\left(|\tfrac{1}{2},\pm\tfrac{1}{2}\rangle;|\tfrac{1}{2},\pm\tfrac{1}{2}\rangle\right)\otimes \left(|\tfrac{1}{2},\pm\tfrac{1}{2}\rangle;|\tfrac{1}{2},\pm\tfrac{1}{2}\rangle\right)=\begin{array}{c}
(\mathrm{singlet};\mathrm{singlet}) +(\mathrm{triplet};\mathrm{singlet})\\
+(\mathrm{singlet};\mathrm{triplet})+(\mathrm{triplet};\mathrm{triplet})
\end{array}
\end{equation}
where $\mathrm{singlet}\equiv|0,0\rangle$ and $\mathrm{triplet}\equiv|1,1\rangle,\,|1,0\rangle,\,|1,-1\rangle$. The total average spin associated with each of these four sets of spin states is $(0\pm0)/2$, $(1\times3\pm0)/(2\times3)$, $(0\pm1\times3)/(2\times3)$, and $(1\times3\pm1\times3)/(2\times3)$, but the middle two are excluded by spin statistics. So, for zero orbital angular momentum, we can expect pseudoscalar and vector $H^\updownarrow_s\overline{H^\updownarrow_t}$. Similarly, for baryons $H^\updownarrow_sH^\updownarrow_tH^\updownarrow_u:=
\mathrm{P}^\updownarrow\Bold{H}_s^+\mathrm{P}^\updownarrow\Bold{H}_t^+\mathrm{P}^\updownarrow\Bold{H}_u^+
+\mathrm{P}^\updownarrow\Bold{H}_s^-\mathrm{P}^\updownarrow\Bold{H}_t^-\mathrm{P}^\updownarrow\Bold{H}_u^-$ (antisymmetry in color indices implied), the average spin combinations turn out to be $\{0,\tfrac{1}{2},1,\tfrac{3}{2},2\}$. In this case, spin statistics rule out the integers and we can expect  $J=\tfrac{1}{2}$ and $J=\tfrac{3}{2}$.

\begin{remark}In both composites the only combinations that lead to acceptable spin statistics derive from iquark pairs in the same spin state. That is, in hadrons,  on average the two components of valence $H_s^\updownarrow$ have aligned spins. As far as the low-energy effective theory based on $SU_S(3)\times U_{EM}(1)$ is concerned then, there is really no material difference between valence quarks and valence $H_s^\updownarrow$ --- they posses equivalent strong charges, extrinsic electric charges, and spin. On the other hand, the EW interaction is based on $\Bold{H}^\pm$. Heuristically speaking, we might say that HC (in hadrons at least) have a split personality because $H^\updownarrow$ want to couple to massless bosons while $\Bold{H}^\pm$ want to couple to massive bosons.
\end{remark}

The mesons and baryons are
composites of $H^\updownarrow_s\overline{H_t^\updownarrow}$ and
$H^\updownarrow_sH^\updownarrow_tH^\updownarrow_u$ respectively where $s,t,u$ label
iquark generation and the up/down arrow denotes either/or. Table \ref{mesons} below contains proposed
assignments for pseudoscalar mesons comprised of the first two generations of composites
$H^\updownarrow_1\overline{H^\updownarrow_1}$,
$H^\updownarrow_1\overline{H^\updownarrow_2}$, and
$H^\updownarrow_2\overline{H^\updownarrow_2}$ (spin content is ignored).

\begin{table}[H]
\centering
\begin{tabular}{||c|c|c|l||}\hline\hline
 HC composite & Iquark composite & Meson & $J^{PC}$ \\ \hline\hline
$\left.\begin{array}{c}
    H_1^\uparrow\overline{H_1^\downarrow}\\
    H_1^\downarrow\overline{H_1^\uparrow}
  \end{array}\right\}$ & $\begin{array}{c}
    h_1^+\overline{\xi_1^0}+\chi_1^0\overline{h_1^-}  \\
    h_1^-\overline{\chi_1^0}+\xi_1^0\overline{h_1^+}
  \end{array}$ &  $\left\{\begin{array}{c}
    \pi^{+} \\
    \pi^{-}
  \end{array}\right.$ & $0^{-}$
\\
&&&\\
 $H_1^\uparrow\overline{H_1^\uparrow}+ H_1^\downarrow\overline{H_1^\downarrow}$
 & $h_1^+\overline{h_1^+}+\chi_1^0\overline{\chi_1^0}+h_1^-\overline{h_1^-}+\xi_1^0\overline{\xi_1^0}$
 &  $\pi^0$ & $0^{-+}$
\\
&&&\\
 $\left.\begin{array}{c}
    H_1^\uparrow\overline{H_2^\downarrow} \\
    H_2^\downarrow\overline{H_1^\uparrow}
  \end{array}\right\}$ & $\begin{array}{c}
    h_1^+\overline{\xi_1^0}+\chi_2^0\overline{h_2^-}  \\
    h_2^-\overline{\chi_2^0}+\xi_1^0\overline{h_1^+}
  \end{array}$ &  $\left\{\begin{array}{c}
    K^{+} \\
    K^{-}
  \end{array}\right.$ & $0^{-}$
\\
&&&\\
 $H_1^\downarrow\overline{H_2^\downarrow}\pm H_2^\downarrow\overline{H_1^\downarrow}$
  & $(h_1^-\overline{h_2^-}+\xi_1^0\overline{\xi_2^0})\pm (h_2^-\overline{h_1^-}+\xi_2^0\overline{\xi_1^0})$
  & $K^{0}_L,\;
    K^0_S$ & $0^{-\pm}$
\\
&&&\\
 $H_1^\uparrow\overline{H_1^\uparrow}+ H_2^\downarrow\overline{H_2^\downarrow}$
 &  $h_1^+\overline{h_1^+}+\chi_1^0\overline{\chi_1^0}+h_2^-\overline{h_2^-}+\xi_2^0\overline{\xi_2^0}$
 &  $\eta$ & $0^{-+}$
\\
&&&\\
 $H_1^\downarrow\overline{H_1^\downarrow}+ H_2^\downarrow\overline{H_2^\downarrow}$
 &  $h_1^-\overline{h_1^-}+\xi_1^0\overline{\xi_1^0}+h_2^-\overline{h_2^-}+\xi_2^0\overline{\xi_2^0}$
 &  $\eta'$ & $0^{-+}$
\\
&&&\\
 $\left.\begin{array}{c}
    H_2^\uparrow\overline{H_1^\downarrow} \\
    H_1^\downarrow\overline{H_2^\uparrow}
  \end{array}\right\}$
  & $\begin{array}{c}
    h_2^+\overline{\xi_2^0}+\chi_1^0\overline{h_1^-}  \\
    h_1^-\overline{\chi_1^0}+\xi_2^0\overline{h_2^+}
  \end{array}$
  &  $\left\{\begin{array}{c}
    D^{+} \\
    D^{-}
  \end{array}\right.$ & $0^{-}$
\\
&&&\\
 $H_1^\uparrow\overline{H_2^\uparrow}\pm H_2^\uparrow\overline{H_1^\uparrow}$
 & $(h_1^+\overline{h_2^+}+\chi_1^0\overline{\chi_2^0})\pm (h_2^+\overline{h_1^+}+\chi_2^0\overline{\chi_1^0})$
 & $ D_+^{0},\;
    D_-^0$ & $0^{-\pm}$
\\
&&&\\
 $\left.\begin{array}{c}
    H_2^\uparrow\overline{H_2^\downarrow} \\
    H_2^\downarrow\overline{H_2^\uparrow}
  \end{array}\right\}$
  & $\begin{array}{c}
    h_2^+\overline{\xi_2^0}+\chi_2^0\overline{h_2^-}  \\
    h_2^-\overline{\chi_2^0}+\xi_2^0\overline{h_2^+}
  \end{array}$
  &  $\left\{\begin{array}{c}
    D_S^{+} \\
    D_S^{-}
  \end{array}\right.$ & $0^{-}$
\\
&&&\\
 $H_2^\uparrow\overline{H_2^\uparrow}+ H_2^\downarrow\overline{H_2^\downarrow}$
 & $h_2^+\overline{h_2^+}+\chi_2^0\overline{\chi_2^0}+h_2^-\overline{h_2^-}+\xi_2^0\overline{\xi_2^0}$
  &  $\eta_c$ & $0^{-+}$
\\
\hline\hline
\end{tabular}
\caption{Proposed HC assignments for selected mesons. Field superscripts denote intrinsic electric charge, subscripts denote iquark generation, and the overbar denotes anti-particle. $J$ is total momentum, $P=(-1)^{L-1}$ is parity, and $C$ indicates $SU_S(3)\times U_{EM}(1)$-conjugation.}
\label{mesons}
\end{table}
As evidenced from the table, it is straightforward to establish the iquark content of the charged mesons, and it is natural to guess that neutral mesons are linear combinations of both $H^\uparrow$ and $H^\downarrow$. With the exception of $\eta$, $\eta'$, and $\eta_c$\footnote{Simple combinatorics suggest the meson assignments in the table, but the HC content can clearly be adjusted to match the standard quark model.} this recipe yields iquark flavor content that matches the quark flavor assignments of the standard quark model if we identify the up-type quark/antiquark composites  with $H^\uparrow_s\overline{H^\uparrow_t}$, down-type composites with $H^\downarrow_s\overline{H^\downarrow_t}$, and mixed-type with $H_s^\updownarrow\overline{H^\updownarrow_t}$. Note that the two mixed-generation neutral combinations give rise to both $CP$-odd and $CP$-even states. It is interesting that  $D_+^0=D^0+\overline{D^0}$ and $D_-^0=D^0-\overline{D^0}$ seem to exhibit very little mass and lifetime asymmetry implying no $CP$ violation\cite{LHCb} despite their $CP$ asymmetry (unlike $K^0_L$ and $K^0_S$).\footnote{Or perhaps $CP$ is violated but the effect is suppressed by the expected relative large mass of $h_2^+$. \emph{Note Added}: Direct $CP$ violation for the $D_+^0,\,D_-^0$ pair has now been observed.\cite{LHCb2}}

Moving on, Table \ref{baryons} contains proposed assignments of selected spin $1/2$
and $3/2$ baryons to HC composites
$H^\updownarrow_1H^\updownarrow_1H^\updownarrow_1$,
$H^\updownarrow_1H^\updownarrow_1H^\updownarrow_2$,
$H^\updownarrow_1H^\updownarrow_2H^\updownarrow_2$, and
$H^\updownarrow_2H^\updownarrow_2H^\updownarrow_2$.

\begin{table}[H]
\centering
\bgroup
\def\arraystretch{1.25}
\begin{tabular}{||c|l|c|c||}\hline\hline
 HC composite & Iquark composite &   Baryon & $J^P$
 \\ \hline\hline
 $H_1^\uparrow H_1^\uparrow H_1^\downarrow$
& $h_1^+h_1^+\xi_1^0+\chi_1^0\chi_1^0 h_1^-$
&  $p$ & $1/2^+$
\\
 $H_1^\uparrow H_1^\downarrow H_1^\downarrow$
& $h_1^+\xi_1^0\xi_1^0+\chi_1^0 h_1^- h_1^-$
&   $n$&$1/2^+$
\\
$H_1^\downarrow H_1^\downarrow H_1^\downarrow$
& $\xi_1^0\xi_1^0 \xi_1^0 +h_1^-h_1^-h_1^-$
&  $\Delta^-$&$3/2^+$
\\
 $H_1^\uparrow H_1^\uparrow H_1^\uparrow$
& $h_1^+h_1^+h_1^++\chi_1^0\chi_1^0 \chi_1^0$
& $\Delta^{++}$&$3/2^+$
\\
 $H_1^\uparrow H_1^\uparrow H_2^\downarrow$
& $h_1^+h_1^+\xi_2^0+\chi_1^0\chi_1^0 h_2^-$
 &  $\Sigma^+$&$1/2^+$
\\\hline
 $H_1^\uparrow H_1^\downarrow H_2^\downarrow$
 & $h_1^+\xi_1^0\xi_2^0+\chi_1^0h_1^- h_2^-$
 &   $\Sigma^0,\,\Lambda^0$&$1/2^+$
\\
 $H_1^\uparrow H_1^\downarrow H_2^\uparrow$
 & $h_1^+\xi_1^0h_2^++\chi_1^0h_1^- \chi_2^0$
 &   $\Sigma_c^+,\,\Lambda_c^+$&$1/2^+$
\\\hline
 $H_1^\downarrow H_1^\downarrow H_2^\downarrow$
 & $\xi_1^0\xi_1^0 \xi_2^0 +h_1^-h_1^-h_2^-$
 &  $\Sigma^-$&$1/2^+$
\\
 $H_1^\uparrow H_2^\downarrow H_2^\downarrow$
 & $h_1^+\xi_2^0\xi_2^0+\chi_1^0h_2^- h_2^-$
 &  $\Xi^0$&$1/2^+$
\\
 $H_1^\downarrow H_2^\downarrow H_2^\downarrow$
 & $\xi_1^0\xi_2^0 \xi_2^0 +h_1^-h_2^-h_2^-$
 &  $\Xi^-$&$1/2^+$
\\
 $H_1^\uparrow H_1^\uparrow H_2^\uparrow$
 & $h_1^+h_1^+h_2^++\chi_1^0\chi_1^0 \chi_2^0$
 &  $\Sigma_c^{++}$&$1/2^+$
\\
 $H_1^\downarrow H_1^\downarrow H_2^\uparrow$
 & $\xi_1^0\xi_1^0 h_2^++h_1^-h_1^-\chi_2^0$
 &  $\Sigma_c^0$&$1/2^+$
\\\hline
 $H_1^\uparrow H_2^\downarrow H_2^\uparrow$
 & $h_1^+\xi_2^0h_2^++\chi_1^0h_2^-\chi_2^0$
 &  ${\Xi'}_c^+,\,\Xi_c^+$&$1/2^+$
\\
 $H_1^\downarrow H_2^\downarrow H_2^\uparrow$
 & $\xi_1^0\xi_2^0 h_2^++h_1^-h_2^-\chi_2^0$
 &  ${\Xi'}_c^0,\,\Xi_c^0$&$1/2^+$
\\\hline
 $H_1^\uparrow H_2^\uparrow H_2^\uparrow$
 & $h_1^+h_2^+h_2^++\chi_1^0\chi_2^0 \chi_2^0$
 & $\Xi_{cc}^{++}$&$1/2^+$
\\
 $H_1^\downarrow H_2^\uparrow H_2^\uparrow$

 & $\xi_2^0h_2^+ h_2^++h_1^-\chi_2^0\chi_2^0$&  $\Xi_{cc}^+$&$1/2^+$
\\
 $H_2^\downarrow H_2^\downarrow H_2^\uparrow$
 & $\xi_2^0\xi_2^0 h_2^++h_2^-h_2^-\chi_2^0$
 &  $\Omega_c^0$&$1/2^+$
\\
 $H_2^\downarrow H_2^\downarrow H_2^\downarrow$
 & $\xi_2^0\xi_2^0 \xi_2^0 +h_2^-h_2^-h_2^-$
 &  $\Omega^-$&$3/2^+$
\\
 $H_2^\downarrow H_2^\uparrow H_2^\uparrow$
 & $\xi_2^0h_2^+h_2^++h_2^-\chi_2^0\chi_2^0$
 &  $\Omega_{cc}^+$&$1/2^+$
\\
 $H_2^\uparrow H_2^\uparrow H_2^\uparrow$
 & $h_2^+h_2^+h_2^++\chi_2^0\chi_2^0 \chi_2^0$
 &   $\Omega_{ccc}^{++}$&$3/2^+$
\\
\hline\hline
\end{tabular}
\egroup
\caption{HC and iquark assignments for selected spin $1/2$ and
$3/2$ baryons.} \label{baryons}
\end{table}
Whether one adheres to color or strong charge as $SU(3)$ quantum numbers, it is easy to check that the iquark composites contribute the requisite (vanishing) strong charges and extrinsic electric charge of the corresponding baryon.

Notice there are four exceptional entries in this table of the form $(H_s^\uparrow H_s^\downarrow)H_t^\updownarrow$ with $s\neq t$. Each is associated with two particles of unequal mass. In the standard quark model, this is attributed to alignment or anti-alignment of isospin. It is noteworthy that the larger mass state primarily decays into its smaller mass partner plus either $\gamma$ (for $\Sigma^0$, ${\Xi'}_c^+$, and ${\Xi'}_c^0$) or $\pi^0$ (for $\Sigma_c^+$). Can the iquark picture explain this?

Recall that $SU_S(3)\times U_{EM}(1)$ sees $H_s^\updownarrow$ like an up/down quark with definite strong charges, extrinsic electric charge, and spin. Consider the case $J=\tfrac{1}{2}$, then  $(H_s^\uparrow H_s^\downarrow)H_t^\updownarrow$ has either $H_s^\uparrow H_s^\downarrow$ or $H_s^\uparrow H_t^\downarrow$ with aligned spins. We propose to attribute the mass splitting  (between the two baryons in each of the four exceptional entries) to two effects: First, when $J=\tfrac{1}{2}$ there must be spin/anti-spin coupling between either the same or different generations. Second, the pair $H_s^\uparrow H_s^\downarrow$ contains the iquark content of $\Bold{H}^+$ and $\Bold{H}^-$ in the same generation. But (with spins aligned) this is precisely what is required to couple to  $A$ and $Z^0$.\footnote{Notice that $Z^0$ is very happy to decay into $\pi^0$.} On the other hand, while $H_s^\uparrow H_t^\downarrow$ also contains the iquark content of $\Bold{H}^+$ and $\Bold{H}^-$, the mixed-generation iquarks can only couple to $W^\pm$ which is a much slower process. The point is, dynamics associated with spin coupling together with two different (same-generation v.s mixed-generation) EW current couplings render two possible states. Since the neutral current interactions have shorter mean lifetimes relative to charged current interactions, we expect a mass splitting between the two states and different decay rates.

\begin{remark}
The combinatorics used to tabulate the hadrons can instead be performed in the context of approximate $SU(4)$ flavor symmetry in the same way that quarks are combined in the standard quark model of hadrons. This would require the assignment of associated quantum numbers to $H^\updownarrow$ and revised HC  content that would mirror the quark model classification scheme given the identifications $H_1^\uparrow\sim u$, $H_1^\downarrow\sim d$, $H_2^\uparrow\sim c$, and $H_2^\downarrow\sim s$. There is certainly precedence in favor of such a scheme. However, (as we discuss in the next subsection) the two approaches make different predictions about the nature of $\Omega_{cb}^0\equiv csb\equiv H_2^\uparrow H_2^\downarrow H_3^\downarrow$: our classification predicts a two-mass state while the standard quark model predicts a single state due to zero isospin. There is not yet definitive particle data so the jury is still out on which classification scheme is correct on this account.
\end{remark}

\subsubsection{Weak phenomenology}
 It is instructive to look at some specific iquark interactions in the particle picture.

\vspace{.1in}

\textbf{Pseudoscalar meson decays}
\begin{itemize}
\item $\pi^0$ decay: Iquark content is $(h_1^+,\,\chi_1^0,\,\overline{h_1^+},\,\overline{\chi_1^0},\,h_1^-,\,\xi_1^0,\,\overline{h_1^-},\,\overline{\xi_1^0})$. Kinematically, $\pi^0$ must decay via $U_{EW}(2)$. As far as the EW interaction is concerned, $\pi^0$ contains neutral currents that couple primarily to $A$ through $h^\pm$: Coupling to $Z^0$ can also occur, but it requires sufficiently localized charged and neutral iquarks and is far weaker. Hence, the primary decay mode in the iquark picture includes annihilation of $\chi^0,\overline{\chi^0}$ and $\xi^0,\overline{\xi^0}$ into gluons via the strong interaction accompanied by $h^\pm,\overline{h^\pm}$ annihilation into $\gamma\gamma$ via the EW interaction. The $\chi^0,\overline{\chi^0}$ and $\xi^0,\overline{\xi^0}$ annihilation obviously proceeds on a much faster time scale.

\item $\eta,\,\eta',\,\eta_c$ decay: Iquark content of $\eta$ for example is $(h_1^+,\,\chi_1^0,\,\overline{h_1^+},\,\overline{\chi_1^0},\,h_2^-,\,\xi_2^0,\,\overline{h_2^-},\,\overline{\xi_2^0})$.
    There are two modes of decay available: The first is the same as that for $\pi^0$ with the valence iquarks contributing to neutral currents coupling primarily to $A$. So the primary EM decay mode and process are the same as for $\pi^0$. The second mode is generation-changing decay $H_2^{\downarrow\uparrow}\rightarrow H_1^{\uparrow\downarrow} +\overline{H_1^{\uparrow\downarrow}}H_1^{\downarrow\uparrow}$ via $W^\pm$. The product valence HC can then combine into various $\pi$ meson combinations yielding for example $\eta\rightarrow \mathrm{n}\pi^0+\mathrm{m}\pi^\pm$. The two decay modes are not mutually exclusive, but decays into n$\pi^0\gamma$ are prohibited by $C$ invariance of EM. However, there are kinematically allowed secondary decay modes of n$\pi^0\gamma\gamma$ and $\pi^+\pi^-\gamma$ as well as intermediate decays into heavier mesons for $\eta'$ and $\eta_c$.

\item $m_{st}^0:=K^0_L,\,K^0_S,\,D^0_+,\,D^0_-$ decay: Iquark content of $K^0$ is $(h_1^-,\,\xi_1^0,\,\overline{h_2^-},\,\overline{\xi_2^0},\,h_2^-,\,\xi_2^0,\,\overline{h_1^-}
    ,\,\overline{\xi_1^0})$. Likewise, for $D^0$ it is $(h_1^+\,\chi_1^0,\,\overline{h_2^+},\,\overline{\chi_2^0},\,h_2^+,\,\chi_2^0,\,\overline{h_1^+}
    ,\,\overline{\chi_1^0})$. Evidently $m_{st}^0$ has just the right  valence iquark content to exhibit generation-changing decay primarily through the process $H_2^{\downarrow\uparrow}\rightarrow H_1^{\uparrow\downarrow}+\overline{H_1^{\uparrow\downarrow}}H_1^{\downarrow\uparrow}$ via $W^\pm$, and the resulting set of valence HC can combine into various collections of mesons and/or leptons. Note that $m_{st}^0$, being a mixed-generation meson by construction, represents both $CP$-even and $CP$-odd states. $CP$-even states can decay into an even number of mesons while the $CP$-odd states decay into an odd number of mesons and/or leptons (to the extent that $CP$ is conserved).

\item $m_{st}^\pm:=\pi^\pm,\,K^\pm,\,D^\pm,\,D_S^\pm$ decay: Iquark content of $\pi^+$ for example  is $({h_1^+},\,\overline{\chi_1^0},\,\overline{h_1^-},\,{\xi_1^0})$. Note that $m_{st}^\pm$ couples to $A$ with total \emph{extrinsic} electric charge $\pm2/3\mp(-1/3)$.  Since $m_{st}^\pm$ iquark weak currents are comprised of \emph{both} $H^\uparrow$ and $H^\downarrow$ constituents, there are potentially two modes available: Scattering of $H^\uparrow,\,H^\downarrow$ into $W^\pm$ and decay of $H_2^{\uparrow\downarrow}$ if present. So, $m_{st}^\pm$ will decay via $W^\pm$ into mesons comprised of both up-type and down-type iquarks and/or leptons. So, for example, the possible leptonic decay modes  are $m_{st}^+\rightarrow \mathrm{mesons}+ l_u^+\nu_v$ and $m_{st}^-\rightarrow \mathrm{mesons}+ l_u^-\overline{\nu_v}$ or $m_{st}^\pm\rightarrow \mathrm{n}\pi^0+\mathrm{m}\pi^\pm$.

\end{itemize}

\vspace{.1in}

\textbf{Baryon EW decays}
\begin{itemize}
\item $n$ decay: The only primary decay mode available is $H_1^\downarrow\rightarrow H_1^\uparrow+\mathrm{leptons}$ via $W^-$. The same decay mode is not kinematically available to $p\equiv H_1^\uparrow H_1^\uparrow H_1^\downarrow$ because the resulting particle $H_1^\uparrow H_1^\uparrow H_1^\uparrow$ requires aligned spins forming a higher energy $J=\tfrac{3}{2}$ state.

\item $b_{(ss)t}:=(H_s^\uparrow H_s^\downarrow)H_t^\updownarrow$ with $s\neq t$ decay. We have already considered these exceptional cases and concluded their peculiar behavior stems from a dynamical interplay between spin coupling and EW current coupling of mixed-generations. If we include the third generation of iquarks, we expect $udb\equiv H_1^\uparrow H_1^\downarrow H_3^\downarrow$ to be manifested as two particles $\Sigma_{b}^0$ and ${\Lambda}_{b}^0$ (as does the standard quark model). However, $csb\equiv H_2^\uparrow H_2^\downarrow H_3^\downarrow$ is also predicted to manifest as two particles ${\Omega'}_{cb}^0$ and $\Omega_{cb}^0$ in the iquark picture even though both states have zero isospin (in which case the standard quark model predicts a single state). The particle data is still inconclusive: the first case seems to hold but neither particle in the second case has been observed.

\item $b_{(st)s}:=(H_s^\uparrow H_t^\downarrow)H_s^\updownarrow$ with $s\neq t$ decay. Here we expect the iquark content from $H_s^\uparrow H_t^\downarrow$ to favor weak decays unless $H_s^\updownarrow$ makes strong decay modes available kinematically. For example $\Sigma^+$ and $\Xi^0$ have $H_1^\uparrow$ and $H_2^\downarrow$ respectively, so our heuristic suggests the primary decay mode is weak. On the other hand, $\Sigma_c^0$ and $\Xi_{cc}^+$ have $H_1^\downarrow$ and $H_2^\uparrow$ so we expect strong decay. These expectations are borne out for the first three, but the $\Xi_{cc}^+$ decay mode has not been established.
\end{itemize}

\subsubsection{Weak $S$-matrix}

We have already argued that the QCD sector of the SM and the n-SM are identical. It remains to establish equivalence\footnote{To be clear; we mean equivalence with the SM as a Yang-Mills QFT not the standard quark model: We have already seen that the iquark picture in the n-SM does not completely agree with the approximate-flavor-symmetry classification scheme of the standard quark model.} between the n-SM and the SM $S$-matrix amplitudes for weak interactions. So it's time to display explicit details regarding the relevant QFT aspects.\footnote{We will follow the notation and conventions of Weinberg\cite{W3} in this subsection.}

To remind, our focus is on the iquark Lagrangian
\begin{eqnarray} \label{lagrangian 2}
\mathcal{L}_{\text{iquark}}&=&i\sum_{s}
 {(\overline{\Bold{H}_{{\mathrm{L}},s}^+}\widetilde{\D}^+
 \Bold{H}^+_{{\mathrm{L}},s}+\overline{\Bold{h}_{{\mathrm{R}},s}^+}\widetilde{\D}^+
 \Bold{h}^+_{{\mathrm{R}},s}+\overline{\Bold{\xi}_{\mathrm{R,s}}^0}\widetilde{\D}\Bold{\xi}^0_{\mathrm{R},s})}\nonumber\\
 &&\hspace{.3in}
 +(\overline{\Bold{H}_{{\mathrm{L}},s}^-}\widetilde{\D}^-
 \Bold{H}^-_{{\mathrm{L}},s}+\overline{\Bold{h}_{{\mathrm{R}},s}^-}\widetilde{\D}^-
 \Bold{h}^-_{{\mathrm{R}},s}+\overline{\Bold{\chi}_{\mathrm{R,s}}^0}\widetilde{\D}\Bold{\chi}^0_{\mathrm{R},s})
\end{eqnarray}
where we have absorbed the scaling constants $\kappa^\pm$ into the covariant derivatives $\widetilde{\D}^\pm:=\kappa^\pm\D^\pm$ which is conceptually appropriate. The $\Bold{H}^+$ field iquark components are
\begin{equation}
h_l^+(x):=(2\pi)^{-3/2}\sum_{\sigma}\int d^3p \, [u_l(\mathbf{p},\sigma,n_{h^+})a(\mathbf{p},\sigma, n_{h^+})e^{ip\cdot x}
+v_l(\mathbf{p},\sigma, n_{h^+})a^\dag(\mathbf{p},\sigma, n^c_{h^+})e^{-ip\cdot x}]
\end{equation}
and
\begin{equation}
\xi_l^0(x):=(2\pi)^{-3/2}\sum_{\sigma}\int d^3p \, [s_l(\mathbf{p},\sigma, n_{\xi^0})a(\mathbf{p},\sigma,n_{\xi^0})e^{ip\cdot x}
+t_l(\mathbf{p},\sigma, n_{\xi^0})a^\dag(\mathbf{p},\sigma, n^c_{\xi^0})e^{-ip\cdot x}]
\end{equation}
where $n_{(\cdot)}$ denotes the $SU_S(3)\times U_{EM}(1)$ quantum numbers and mass of the indicated iquark. There are analogous fields for $\Bold{H}^-$, and they all combine to give the up/down iquark fields
\begin{eqnarray}
H_l^\uparrow(x)&:=&(2\pi)^{-3/2}\sum_{\sigma}\int d^3p \,\left\{
[u_l(\mathbf{p},\sigma,n_{h^+})a(\mathbf{p},\sigma, n_{h^+})+s_l(\mathbf{p},\sigma,n_{\chi^0})a(\mathbf{p},\sigma, n_{\chi^0})]e^{ip\cdot x}\right.\notag\\
&&
\hspace{1in}+\left.[v_l(\mathbf{p},\sigma,n_{h^+})a^\dag(\mathbf{p},\sigma, n_{h^+})+t_l(\mathbf{p},\sigma,n_{\chi^0})a^\dag(\mathbf{p},\sigma, n_{\chi^0})]e^{-ip\cdot x}
\right\}\notag\\
\end{eqnarray}
and
\begin{eqnarray}
H_l^\downarrow(x)&:=&(2\pi)^{-3/2}\sum_{\sigma}\int d^3p \,\left\{
[u_l(\mathbf{p},\sigma,n_{h^-})a(\mathbf{p},\sigma, n_{h^-})+s_l(\mathbf{p},\sigma,n_{\xi^0})a(\mathbf{p},\sigma, n_{\xi^0})]e^{ip\cdot x}\right.\nonumber\\
&&
\hspace{1in}+\left.[v_l(\mathbf{p},\sigma,n_{h^-})a^\dag(\mathbf{p},\sigma, n_{h^-})+t_l(\mathbf{p},\sigma,n_{\xi^0})a^\dag(\mathbf{p},\sigma, n_{\xi^0})]e^{-ip\cdot x}
\right\}\;.\notag\\
\end{eqnarray}

With the weak interaction turned off, there is no distinction between $\overline{\Bold{H}_{{\mathrm{L}},s}^+}\widetilde{\D}^+
 \Bold{H}^+_{{\mathrm{L}},s}$ and $\overline{\Bold{H}_{{\mathrm{L}},s}^-}\widetilde{\D}^-
 \Bold{H}^-_{{\mathrm{L}},s}$ so $\mathcal{L}_{\text{iquark}}$ reduces to the fermion kinetic term $\mathcal{L}_{\text{quark}}$ of QCD in this limit. Consequently iquarks and quarks share identical free-field propagators in their respective perturbative QFTs. Now turn on the weak interaction and compare the EW interaction Feynman rules for $H_l^{\uparrow\downarrow}$ v.s. up/down quarks $U,D$:

\vspace{.4in}
\begin{tabular}{ccc}
\underline{\textbf{n-SM iquark/neutral-boson}} & & \underline{\textbf{SM quark/neutral-boson}} \\
\vspace{.2in}\\
\begin{tikzpicture}[line width=1.0 pt, scale=.8]
	
			\draw[fermion] (-140:1)--(0,0);
			\draw[fermion] (140:1)--(0,0);
			\draw[provector] (0:1)--(0,0);
			\node at (-140:1.75) {$(h^+;\chi^0)$};
			\node at (140:1.75) {$(\overline{h^+};\overline{\chi^0})$};
			\node at (0:3.5) {$A\;\sim\;i(+\frac{2}{3}e\,;-\frac{1}{3}0)\gamma^\mu$};	
\end{tikzpicture}

& \hspace{.1in} &
\begin{tikzpicture}[line width=1.0 pt, scale=.8]
	
			\draw[fermion] (-140:1)--(0,0);
			\draw[fermion] (140:1)--(0,0);
			\draw[provector] (0:1)--(0,0);
			\node at (-140:1.75) {$U$};
			\node at (140:1.75) {$\overline{U}$};
			\node at (0:2.5) {$A\;\sim\; i\frac{2}{3}e\gamma^{\mu}$};	
\end{tikzpicture}
\\
\\
\begin{tikzpicture}[line width=1.0 pt, scale=.8]
	
			\draw[fermion] (-140:1)--(0,0);
			\draw[fermion] (140:1)--(0,0);
			\draw[provector] (0:1)--(0,0);
			\node at (-140:1.75) {$(h^-;\xi^0)$};
			\node at (140:1.75) {$(\overline{h^-};\overline{\xi^0})$};			
            \node at (0:3.5) {$A\;\sim\;i(-\frac{1}{3}e\,;+\frac{2}{3}0)\gamma^\mu$};	
\end{tikzpicture}

& \hspace{.1in} &
\begin{tikzpicture}[line width=1.0 pt, scale=.8]
	
			\draw[fermion] (-140:1)--(0,0);
			\draw[fermion] (140:1)--(0,0);
			\draw[provector] (0:1)--(0,0);
			\node at (-140:1.75) {$D$};
			\node at (140:1.75) {$\overline{D}$};
			\node at (0:2.6) {$A\;\sim\; -i\frac{1}{3}e\gamma^{\mu}$};	
\end{tikzpicture}
\end{tabular}

\begin{tabular}{ccc}
&\vspace{0in}&\\
\begin{tikzpicture}[line width=1.0 pt, scale=.7]
	
			\draw[fermion] (-140:1)--(0,0);
			\draw[fermion] (140:1)--(0,0);
			\draw[provector] (0:1)--(0,0);
			\node at (-140:1.75) {$(h^+;\chi^0)$};
			\node at (140:1.75) {$(\overline{h^+};\overline{\chi^0})$};
			\node at (0:5.1) {$Z^0\;\sim\; i(+\frac{2}{3}F(\theta)g_{Z^0}\,;-\frac{1}{3}g_{Z^0})\gamma^\mu$};	
\end{tikzpicture}
& &\hspace{-.2in}
\begin{tikzpicture}[line width=1.0 pt, scale=.7]
	
			\draw[fermion] (-140:1)--(0,0);
			\draw[fermion] (140:1)--(0,0);
			\draw[provector] (0:1)--(0,0);
			\node at (-140:1.75) {$U$};
			\node at (140:1.75) {$\overline{U}$};
			\node at (0:4.4) {$Z^0\;\sim\;i(\frac{2}{3}F(\theta)-\frac{1}{3})g_{Z^0}\gamma^{\mu}$};	
\end{tikzpicture}
\\
\\
\begin{tikzpicture}[line width=1.0 pt, scale=.7]
	
			\draw[fermion] (-140:1)--(0,0);
			\draw[fermion] (140:1)--(0,0);
			\draw[provector] (0:1)--(0,0);
			\node at (-140:1.75) {$(h^-;\xi^0)$};
			\node at (140:1.75) {$(\overline{h^-};\overline{\xi^0})$};
			\node at (0:5.1) {$Z^0\;\sim\;
i(-\frac{1}{3}F(\theta)g_{Z^0}\,;+ \frac{2}{3}g_{Z^0})\gamma^\mu$};	
\end{tikzpicture}

& &\hspace{-.2in}

\begin{tikzpicture}[line width=1.0 pt, scale=.7]
	
			\draw[fermion] (-140:1)--(0,0);
			\draw[fermion] (140:1)--(0,0);
			\draw[provector] (0:1)--(0,0);
			\node at (-140:1.75) {$D$};
			\node at (140:1.75) {$\overline{D}$};
			\node at (0:4.4) {$Z^0\;\sim\; i(\frac{2}{3}-\frac{1}{3}F(\theta))g_{Z^0}\gamma^{\mu}$};	
\end{tikzpicture}
\end{tabular}

\vspace{.3in}

\noindent where $F(\theta_W):=2\sin^2(\theta_W)-1$, and the coupling constant is $g_{Z^0}:=e/(2\cos\theta_W\sin\theta_W)$. The left column actually represents two diagrams; one for each iquark in the ordered pair $(\cdot\,;\,\cdot)$.

Similarly, the charged couplings are

\vspace{.4in}

\begin{tabular}{ccc}
\underline{\textbf{n-SM iquark/charged-boson}} & & \underline{\textbf{SM quark/charged-boson}} \\
\vspace{.2in}\\
\begin{tikzpicture}[line width=1.0 pt, scale=.8]
	
			\draw[fermion] (-140:1)--(0,0);
			\draw[fermion] (140:1)--(0,0);
			\draw[provector] (0:1)--(0,0);
			\node at (-140:1.75) {$(\xi^0;h^-)$};
			\node at (140:1.75) {$(\overline{h^+};\overline{\chi^0})$};
			\node at (0:5) {$W^-\;\sim\; i(+\frac{2}{3}V^+g_W\,; -\frac{1}{3}{V^-}^\dag g_W)\gamma^\mu$};	
\end{tikzpicture}
&  &\hspace{-.2in}
\begin{tikzpicture}[line width=1.0 pt, scale=.8]
	
			\draw[fermion] (-140:1)--(0,0);
			\draw[fermion] (140:1)--(0,0);
			\draw[provector] (0:1)--(0,0);
			\node at (-140:1.75) {$D$};
			\node at (140:1.75) {$\overline{U}$};
			\node at (0:3) {$W^-\;\sim\; iVg_W\gamma^{\mu}$};	
\end{tikzpicture}
\\
\\
\begin{tikzpicture}[line width=1.0 pt, scale=.8]
	
			\draw[fermion] (-140:1)--(0,0);
			\draw[fermion] (140:1)--(0,0);
			\draw[provector] (0:1)--(0,0);
			\node at (-140:1.75) {$(h^+;\chi^0)$};
			\node at (140:1.75) {$(\overline{\xi^0};\overline{h^-})$};
			\node at (0:5) {$W^+\;\sim\; i(+\frac{2}{3}{V^+}^\dag g_W\, ; -\frac{1}{3}V^- g_W)\gamma^\mu$};	
\end{tikzpicture}
&  &\hspace{-.2in}
\begin{tikzpicture}[line width=1.0 pt, scale=.8]
	
			\draw[fermion] (-140:1)--(0,0);
			\draw[fermion] (140:1)--(0,0);
			\draw[provector] (0:1)--(0,0);
			\node at (-140:1.75) {$U$};
			\node at (140:1.75) {$\overline{D}$};
			\node at (0:3) {$W^+\;\sim\; iV^\dag g_W\gamma^{\mu}$};	
\end{tikzpicture}
\end{tabular}

\vspace{.3in}

\noindent where $g_W=e/(\sqrt{2}\sin\theta_W)$.

As previously remarked, each up/down quark coupling is identified with the sum of two corresponding up/down iquark couplings. Given this observation, our aim is to show that, up to a re-definition of elementary-particle content, the n-SM and SM predict identical $S$-matrix amplitudes for EW interactions when $V^+=-{V^-}^\dag\equiv V$.

Expanded in terms of individual iquarks, the n-SM interaction Hamiltonian density is of the form $\mathcal{H}_{\mathrm{n-SM}}(x)=\sum_{klm}\not\!\!\phi_k(x)g_{klm}\psi_l^\dag(x)\psi_m(x)$ where $g_{klm}$ are coupling strengths encoded in (\ref{representation}) and (\ref{lagrangian 2}), $\phi_k(x)$ represents EW gauge bosons and $\psi_l(x)$ represents iquarks. The subscripts indicate Lorentz indices, particle type, and particle generation when relevant. Notice that $g_{klm}$ vanishes for any $lm$ that mix $\pm$ iquark fields. In terms of up/down iquarks,
\begin{eqnarray}
\mathcal{H}_{\mathrm{n-SM}}(x)
&=&\sum_{{k^0}lm}\not\!\!\phi_{k^0}(x)\left({g}^\uparrow_{k^0lm}
{({{\psi}_l^\uparrow}(x))^\dag}{\psi}^\uparrow_m(x)
+{g}^\downarrow_{k^0lm}{({{\psi}_l^\downarrow}(x))}^\dag{\psi}^\downarrow_m(x)\right)\notag\\
&&+\sum_{lm}\left(\not\!\phi_{k^-}(x){g}^-_{k^-lm}{({{\psi}_l^\uparrow}(x))^\dag}{\psi}^\downarrow_m(x)
+ \not\!\phi_{k^+}(x){g}^+_{k^+lm}{({{\psi}_l^\downarrow}(x))^\dag}{\psi}^\uparrow_m(x)\right)\;.
\end{eqnarray}

Likewise, assuming equivalent gauge fixing, the SM counterpart has the same form $\mathcal{H}_{\mathrm{SM}}(x)=\sum_{klm}
\not\!\phi_k(x)\widetilde{g}_{klm}\widetilde{\psi}_l^\dag(x)\widetilde{\psi}_m(x)$ where now $\widetilde{g}_{klm}$ are the usual SM coupling constants and $\widetilde{\psi}_l(x)$ represents quark fields. In terms of up/down quarks,
\begin{eqnarray}
\mathcal{H}_{\mathrm{SM}}(x)&=&\sum_{{k^0}lm}\not\!\phi_{k^0}(x)\left(\widetilde{g}^\uparrow_{k^0lm}
{U_l}^\dag(x)U_m(x)
+\widetilde{g}^\downarrow_{k^0lm}D_l^\dag(x)D_m(x)\right)\notag\\
&&+\sum_{lm}\left(\not\!\phi_{k^-}(x)\widetilde{g}_{k^-lm}U_l^\dag(x)D_m(x)
+\not\!\phi_{k^+}(x)\widetilde{g}_{k^+lm}D_l^\dag(x)U_m(x)\right)\;.
\end{eqnarray}

The two theories are congruent in the sense that matrix elements between their respective  hadronic composites are identical. For example, consider $\pi^+\equiv H_1^\uparrow \overline{H_1^\downarrow}=h_1^+\overline{\xi_1^0}+\chi_1^0\overline{h_1^-}\equiv u\overline{d}$. Then
\begin{eqnarray}\label{congruence}
\left( H_1^\uparrow \overline{H_1^\downarrow},\mathcal{H}_{\mathrm{n-SM}}(x)H_1^\uparrow \overline{H_1^\downarrow}\right)
 &=&ig_W\not\!\!W^+(x)\left[\tfrac{2}{3}{V^+}^\dag\left( h_1^+\overline{\xi_1^0},h_1^+\overline{\xi_1^0}\right)
 +\tfrac{1}{3}V^-\left( \chi_1^0\overline{h_1^-},\chi_1^0\overline{h_1^-}\right)\right]\notag\\
&=&ig_W\not\!\!W^+(x)\left[\tfrac{2}{3}{V^+}^\dag
 +\tfrac{1}{3}V^-\right]=ig_W\not\!\!W^+(x)V^\dag
\end{eqnarray}
where we have imposed $V^+=-{V^-}^\dag\equiv V$.
On the other hand,
\begin{equation}
\left( u\overline{d},\mathcal{H}_{\mathrm{SM}}(x)u\overline{d}\right)=ig_W\not\!\!W^+(x)V^\dag\;.
\end{equation}
Referring to the Feynman diagrams, it is not difficult to see that the matrix elements between any two hadron states  are identical:
\begin{equation}
\big(\mathrm{hadron}',\mathcal{H}_{\mathrm{n-SM}}(x)\mathrm{hadron}\big)
\circeq\big(\widetilde{\mathrm{hadron}}',\mathcal{H}_{\mathrm{SM}}(x)\widetilde{\mathrm{hadron}}\big)\;.
\end{equation}
We are using the term `identical' and the symbol $\circeq$ because each side contains spinor functions that carry iquark/quark labels. So mathematically, the two sides can be distinguished; but since no iquark/quark has been observed they are physically indistinguishable.

More generally, for matrix elements of a state $\Phi_\alpha$ comprised of leptons, mesons, and baryons labeled in either theory
\begin{eqnarray}
\Phi_\alpha(\psi^\updownarrow_{l})
&:=&\mathrm{leptons}+\sum_{l m}m^\updownarrow_{l m}\psi^\updownarrow_{l} \overline{\psi^\updownarrow_{m}}+\sum_{lmn}b^\updownarrow_{lmn}\psi^\updownarrow_{l} \psi^\updownarrow_{m} \psi^\updownarrow_{n}\notag\\
&& \hspace{-.75in}\| \notag\\
\Phi_\alpha(\widetilde{\psi}^\updownarrow_{l} )
&:=&\mathrm{leptons}+\sum_{l m}\widetilde{m}^\updownarrow_{l m}\widetilde{\psi}^\updownarrow_{l} \overline{\widetilde{\psi}^\updownarrow_{l}}+\sum_{lmn}\widetilde{b}^\updownarrow_{lmn}\widetilde{\psi}^\updownarrow_{l} \widetilde{\psi}^\updownarrow_{m} \widetilde{\psi}^\updownarrow_{n}\;\;,
\end{eqnarray}
we have two physically indistinguishable descriptions of the $S$-matrix $\left(\Phi_\beta,S\Phi_\alpha\right)=:S_{\alpha\beta}$ because
\begin{eqnarray}\label{relation}
\left(\Phi_\beta({\psi}^\updownarrow_{l}), T\{\mathcal{H}_{\mathrm{n-SM}}(x_1)\cdots \mathcal{H}_{\mathrm{n-SM}}(x_N)\}\Phi_\alpha(\psi^\updownarrow_{l})\right)\notag\\
&&\hspace{-1.25in}\circeq\left( \Phi_\beta(\widetilde{\psi}^\updownarrow_{l}),T\{\mathcal{H}_{\mathrm{SM}}(x_1)\cdots \mathcal{H}_{\mathrm{SM}}(x_N)\}\Phi_\alpha(\widetilde{\psi}^\updownarrow_{l} )\right)\;.
\end{eqnarray}
Relation (\ref{relation}) follows from straightforward induction on $N$,  inserting complete sets of states appropriately, and using $\big( \Phi_\alpha({\psi}^\updownarrow_{l}),\mathcal{H}_{\mathrm{n-SM}}(x)\Phi_\alpha(\psi^\updownarrow_{l})\big)
 \circeq\big(\Phi_\alpha(\widetilde{\psi}^\updownarrow_{l}),\mathcal{H}_{\mathrm{SM}}(x)\Phi_\alpha(\widetilde{\psi}^\updownarrow_{l} )\big)$.

 Since individual iquarks/quarks are not observed, $|S_{\alpha\beta}|^2$  includes an average over the associated spins. This eliminates the spinors that carry iquark/quark labels, and everything (sans leptons) boils down to momenta, masses, and electric charges of $\Phi_\alpha$ and $\Phi_\beta$ --- no direct link to either ${\psi}^\updownarrow_{l}$ or $\widetilde{\psi}^\updownarrow_{l}$ remains. For states comprised of leptons and hadrons then, \emph{the $S$-matrix for EW transitions in the n-SM and SM are physically indistinguishable and they yield equivalent transition rates and cross sections}. This is not to say that parametrized, model-dependent predictions will necessarily agree: As we have already mentioned, PDFs will certainly differ for charged lepton--hadron DIS phenomenology.

\section{Representation transmutation}\label{trans-representation}
It is noteworthy that the SM and the n-SM have different gauge group and elementary particle content, and yet they make equivalent experimental predictions. The virtue of the n-SM is that it offers new avenues and suggestions for speculative modifications.

It is believed that pure QCD has the potential to exhibit Coulomb, confining, and Higgs phases --- the confining phase being manifest at typical terrestrial energies. With matter included, the phases (as a function of matter field chemical potential and temperature) appear to include the hadronic phase, a quark/gluon plasma phase, a quark liquid phase, and a CFL superfluid phase.

With this backdrop, a notable feature of the n-SM is the equivalent EW representation of iquarks and leptons. Recall this is a direct consequence of a \emph{product} symmetry group. One wonders if the product structure exerts influence elsewhere.

We propose that the product group structure can lead to a phase that does not belong to one of the phase classes described above. Our hypothesis is based on the observation that  a trace over the $\mathfrak{su}_S(3)$ components of the covariant derivative leads to transmuted representations $(\bs{3},\bs{2})_L,(\bs{3},{\mathbf{2^c}})_L\rightarrow(\bs{1},\bs{2})_L,(\bs{1},{\mathbf{2^c}})_L$ and $({\bs{3}},\bs{1}^\pm)_R,({\bs{3}},\bs{1}^0)_R
\rightarrow({\bs{1}},\bs{1}^\pm)_R,({\bs{1}},\bs{1}^0)_R$ (and their $SU_S(3)$ conjugates). In as much as $\kappa^++\kappa^-=1$, such a representation transmutation would look like a phase change from iquarks to leptons --- without $SU_S(3)$ symmetry breaking. In effect, QCD is posited to have a leptonic phase. We will call this hypothetical transition from hadronic phase  to leptonic phase ``trans-representation'' for short.

But what mechanism could possibly trigger trans-representation? One idea is to view the gauge bosons of the product group $SU_S(3)\times U_{EW}(2)$  as a combined system/environment bridged by their mutual coupling to matter fields. Then one could imagine that the EW gauge--matter interaction might see the strong gauge--matter interaction as a source of decoherence. And, under suitable position/momentum/particle-content conditions, the decoherence might accumulate to the point of a phase transition as described. One can even imagine the influence goes the other way. Perhaps at the same or different energy scale the strong gauge--matter interaction sees the EW gauge--matter interaction as an environment. The same representation reduction might occur for $SU_S(3)$, and the eventual EW-energy-scale iquark and lepton content would ultimately spring from $(\bs{3},\bs{2})\oplus(\bs{3},{\mathbf{2^c}})$.

To be more explicit, assume the domain  of the Lagrangian (\ref{lagrangian})
has been restricted to the physical state space which is endowed with a Hilbert structure. Then, since the Lagrangian is a closed and symmetric form, there exist  unique self-adjoint operators that represent the covariant derivatives (\ref{covariant derivative}). So we can go over to an operator picture.

At the beginning of the electroweak epoch, we assume matter fields in representation $(\bs{3},\bs{2})\oplus(\bs{3},{\mathbf{2^c}})$ and boson states $|\phi_{SU(3)}\rangle\otimes|\phi_{U(2)}\rangle\in H_{SU(3)}\otimes H_{U(2)}\subset H_{\mathrm{gauge}}\oplus H_{\mathrm{matter}}=:H$ where $H$ is the total \emph{physical} Hilbert space.

Of course the $SU(3)$ and $U(2)$ gauge bosons do not interact directly. However, they do share common matter fields (including Higgs), and over a sufficiently coarse-grained phase space \emph{some} of the matter--gauge interactions can be viewed as boson--boson scattering (viz. diagrams with internal matter loops). Adopting this picture, we will assume it can be described by a total scattering operator that can be decomposed as
\begin{equation}
\mathbf{S}_{\mathrm{Tot}}=\mathbf{S}_0+\mathbf{S}_1+\mathbf{S}_2+\mathbf{S}_3
\end{equation}
where $\mathbf{S}_0$ is diagonal in the Cartan basis of both $SU(3)$ and $U(2)$ (so it looks like segregated gauge boson scattering), $\mathbf{S}_1$ (resp. $\mathbf{S}_2$) is diagonal in the Cartan basis of $SU(3)$  (resp. $U(2)$), and $\mathbf{S}_3$ is not diagonal in either. These individual components would presumably only be manifest in various subspaces of the total phase space.

For example, $\mathbf{S}_1$ represents an operator corresponding to a scattering event that does not mix the $SU(3)$ Cartan basis states. Then $\mathbf{S}_1$ has the (assumed) form
\begin{equation}
\mathbf{S}_1:=\sum_\alpha|\alpha\rangle\langle \alpha|\otimes\mathbf{S}_\alpha
\end{equation}
where $|\alpha\rangle$ represents the Cartan basis in $\mathfrak{su}_S(3)$ and $\mathbf{S}_\alpha$ are scattering operators representing the influence of gauge boson $|\alpha\rangle$ coupling to $|\phi_{U(2)}\rangle$ via  virtual matter fields. Likewise,
\begin{eqnarray}
\mathbf{S}_0&:=&\sum_{\alpha,\sigma}|\alpha\rangle\langle \alpha|\otimes|\sigma\rangle\langle \sigma|\notag\\
\mathbf{S}_2&:=&\sum_\sigma\mathbf{S}_\sigma\otimes|\sigma\rangle\langle \sigma|\notag\\
\mathbf{S}_3&:=&\sum_{\alpha,\sigma}\mathbf{S}_\sigma\otimes\mathbf{S}_\alpha
\end{eqnarray}

Consider a single scattering event governed by $\mathbf{S}_1$. On a time scale much larger than the interaction time, the initial gauge field joint density is $\rho_{SU(3)}\otimes\rho_{U(2)}$. The final reduced density will then be
\begin{eqnarray}\label{joint density}
\rho^R&=&\mathbf{S}_1\left(\rho_{SU(3)}\otimes\rho_{U(2)}\right)\mathbf{S}_1^\dag\notag\\
&=&\sum_{\alpha,\beta}p_\alpha p_\beta|\alpha\rangle \langle \alpha|\rho_{SU(3)}|\beta\rangle\langle \beta|\otimes\rho_{U(2)}^{\mathbf{S}_{\alpha,\beta}}
\end{eqnarray}
where $p_\alpha,\,p_\beta$ are probabilities and $\rho_{U(2)}^{\mathbf{S}_{\alpha,\beta}}:=\mathbf{S}_\alpha\,\rho_{U(2)}\,\mathbf{S}_\beta^\dag$. The $SU(3)$ contribution to the reduced final density matrix is therefore
\begin{eqnarray}
\rho^R_{SU(3)}=
\mathrm{tr}_{U(2)}(\rho^R)&=&\sum_{\alpha,\beta}p_\alpha p_\beta\langle \alpha|\rho_{SU(3)}|\beta\rangle\left[\mathrm{tr}_{U(2)}(\rho_{U(2)}^{\mathbf{S}_{\alpha,\beta}})\right]
|\alpha\rangle \langle \beta|\;.
\end{eqnarray}

To the extent that $\left[\mathrm{tr}_{U(2)}(\rho_{U(2)}^{\mathbf{S}_{\alpha,\beta}})\right]$ diminishes $|\rho^R_{SU(3)}|$, the off-diagonal matrix elements become `decoherent' in the sense that superpositions of $SU(3)$ gauge bosons that might be induced by the collision can become highly suppressed. On the other hand, for $\alpha=\beta$ the right-hand side of (\ref{joint density}) collapses to $\rho_{SU(3)} \left[\mathrm{tr}_{U(2)}(\rho_{U(2)})\right]$ since $\mathbf{S}_1$ is unitary. Together, these indicate the canonical status of gauge boson states associated with the $SU(3)$ Cartan basis are not disrupted by the collisions encoded in $\rho^R_{SU(3)}$ and the diagonal of $\rho^R$. This is crucial for on-shell iquarks trying to couple to gluons. Recall that \emph{iquarks are eigenfields relative to the Cartan basis}: they can't exchange charge with a  statistical ensemble of gluon states (which would carry an associated mixture of charges).  So, after the scattering, on-shell iquarks in the defining representation can only effectively couple to $\rho^R_{SU(3)}$ and the diagonal part of $\rho^R$.

The reduced final density can alternatively be expressed as
\begin{eqnarray}
\rho^R_{SU(3)}&=&\sum_{\sigma,\varepsilon}p_\sigma \langle\varepsilon|\mathbf{S}_1|\sigma\rangle\,\rho_{SU(3)}\,\langle\sigma|\mathbf{S}_1^\dag|\varepsilon\rangle\notag\\
&=:&\sum_{\sigma,\varepsilon}p_\sigma\mathbf{S}_1(\sigma,\varepsilon)
\,\rho_{SU(3)}\,{\mathbf{S}_1}(\sigma,\varepsilon)^\dag
\end{eqnarray}
where $\mathbf{S}_1(\sigma,\varepsilon)=\sum_{\alpha}|\alpha\rangle\langle \alpha|\,\langle\sigma|\mathbf{S}_\alpha|\varepsilon\rangle$. Unitary $\mathbf{S}_1$ implies
\begin{equation}
\sum_{\sigma,\varepsilon}p_\sigma\mathbf{S}_1(\sigma,\varepsilon)\,
\mathbb{I}_{SU(3)}\,{\mathbf{S}_1}(\sigma,\varepsilon)^\dag=1\;,
\end{equation}
so $\mathbf{S}_1(\sigma,\varepsilon)$ represents an effective scattering operator on $H_{SU(3)}$. We want to associate this operator with the covariant derivatives acting on the $(\mathbf{3},\mathbf{1}^\pm)$ and $(\mathbf{3},\mathbf{1}^0)$ matter fields. Reversing the roles of $SU(3)$ and $U(2)$,  similar conclusions apply for $\rho^R_{U(2)}$. The effective scattering operators $\mathbf{S}_2(\alpha,\beta)$ on $H_{U(2)}$ would lead to the lepton covariant derivative for $(\mathbf{1},\mathbf{2})$. Finally, $\mathbf{S}_3$ would yield the lepton covariant derivatives for $(\mathbf{1},\mathbf{1}^\pm)$ and $(\mathbf{1},\mathbf{1}^0)$.

\begin{remark}
A reasonable guess of how this would work is that meson production along with $\mathbf{S}_\alpha$ would lead to
$(\bs{3},\bs{2})\rightarrow (\bs{3},\bs{2})_L,\,(\bs{3},\bs{1}^+)_R,\,(\bs{3},\bs{1}^0)_R$ and $(\bs{3},{\mathbf{2^c}})\rightarrow (\bs{3},{\mathbf{2^c}})_L,\,(\bs{3},\bs{1}^-)_R,\,(\bs{3},\bs{1}^0)_R$. Likewise for $(\overline{\bs{3}},\overline{\bs{2}})$ and $(\overline{\bs{3}},\overline{{\mathbf{2^c}}})$. Either before, during, or after this would be accompanied by $\mathbf{S}_\sigma$ induced transitions $(\bs{3},\bs{2})\rightarrow (\bs{1},\bs{2})$ and $(\bs{3},\bs{1}^\pm),\,(\bs{3},\bs{1}^0)\rightarrow (\bs{1},\bs{1}^\pm),\,(\bs{1},\bs{1}^0)$. One can verify that inserting these lepton fields into the iquark Lagrangian and using $\kappa^++\kappa^-=1$ yields the usual lepton Lagrangian of the SM. A more detailed accounting of this is included in appendix \emph{A}.

We know the matter field representations supposedly induced by $\mathbf{S}_1$ and $\mathbf{S}_3$. At energy scales above $Q_{EW}$, we might expect $\mathbf{S}_0$ and $\mathbf{S}_2$ induced states to contribute to the Lagrangian according to \emph{(\ref{S0})} and \emph{(\ref{S2})}. Ostensibly, such states would not survive the decoherence and/or electroweak symmetry breaking.

One might guess that $\mathbf{S}_\sigma$ and $\mathbf{S}_\alpha$ are the origin of generation mixing and chiral asymmetry. If so and  assuming a massless pair $(\bs{3},\bs{2}),(\bs{3},{\mathbf{2^c}})$,  there would be an \textbf{approximate} global $\left(U(2)_L\times U(2)_R\right)\times \mathbb{Z}_2$ outer automorphism symmetry in $\mathbf{S}_0$ that would be spontaneously broken via $\mathbf{S}_\alpha$ to an approximate global $SU_V(2)\times U_V(1)$ in $\mathbf{S}_1$ and $\mathbf{S}_3$ --- resulting in pseudo-Goldstone bosons. A possible connection with Higgs\emph{\cite{G2,CO}} merits further investigation of this point.

\end{remark}

Evidently $B+L$ is conserved across all energy scales in this picture. But, between the hypothetical energy scale of $SU(3)\times U(2)$ and the electroweak SSB scale,  baryon and lepton numbers may not be separately conserved because iquarks and leptons can be transmuted. This is potentially problematic for proton decay.

\section{Summary and outlook}
The primary purpose of this paper is to explore some foundational aspects of the SM under the replacement $SU_I\times U_Y(1)\rightarrow \newew$. The replacement is physically well motivated: Among other things, it renders faithful matter field representations and implies the Gell-Mann/Nishijima relation $Q\propto I+1/2Y$.

Due to the semi-direct product structure of $\newew$ and the group-product structure of $SU_S(3)\times \newew$, we argue that the electroweak matter field degrees of freedom, which we call iquarks, furnish two inequivalent defining representations. The structure of the Lie algebra $\mathfrak{u}(2)$ leads directly to the physical basis associated with observed electroweak bosons without (and in fact precluding) introducing isospin and hypercharge and subsequently positing a change of basis precipitated by SSB. Including both iquark representations in the Lagrangian density and fixing their relative couplings by imposing anomaly cancellation leads to strong and electroweak currents that agree with the SM --- despite the fact that iquarks possess integer electric charge --- as long as quarks/iquarks cannot be observed. It is natural to identify gauge boson and matter field quantum numbers with the (re-scaled) Lie algebra root system. This leads to the conclusion that six charged gluons exchange \emph{two} strong charges with each other and with iquarks, and two neutral gluons mediate neutral iquark currents. Evidently the neutral gluons experience a different hadronic confining potential, and their possible role in nuclear bonding is worth investigating.

The Lagrangian density comprised of the two iquark contributions is invariant under an approximate global $(U(2)_L\times U(2)_R)\times\mathbb{Z}_2$ symmetry that reduces to $SU(2)_V\times U(1)_V$ under chiral symmetry breaking. The consequent pseudo-Goldstone boson can be identified as a Higgs-like boson. It is known that such pseudo-Goldstone bosons are exempt from the hierarchy problem. It would be interesting to determine if this one is consistent with LHC phenomenology. We leave it as an open question.

Since $SU_S(3)\times U_{EM}(1)$ iquark currents are identical to SM quark currents (assuming confinement),  n-SM and SM make equivalent QCD and EM predictions. Looking at electroweak phenomenology, the iquark model explains the split-mass level of mixed-generation baryons without invoking approximate flavor symmetry as practiced in the standard quark model. We also prove that the spin-averaged $S$-matrix for n-SM scattering agrees with the spin-averaged SM $S$-matrix. However, parton distributions and structure functions will differ, so parametrized models could conceivably make different predictions. Any differences coming from spin/charge coupling effects are expected to be small due to the presumed small mass of the neutral iquarks.

The $U(2)$ modifications leading to the n-SM motivate a more substantial departure: They suggest the idea of representation transmutation. Representation transmutation is the notion that similarities between the iquark and lepton representations stem from a decoherence battle waged between $SU_S(3)$ and $\newew$. Presumably, there are regions in field phase space where various matter field representations are manifest. Representation transmutation offers a tool to study the connection (if one exists) between the CKM and PMNS generation-mixing matrices.

\appendix
\section{Trans-representation and anomaly cancellation}
 Assume the starting symmetry group is $ SU_S(3)\times U_{EW}(2)$. We propose the matter field content consists of $(\bs{3},\bs{2})$, $(\bs{3},{\mathbf{2^c}})$, $(\overline{\bs{3}},\overline{\bs{2}})$, and $(\overline{\bs{3}},\overline{{\mathbf{2^c}}})$. Suppose, then, the $U_{EW}(2)$ environment induces trans-representation $(\bs{3},\bs{2})\rightarrow (\bs{3},\bs{2})_L\oplus (\bs{3},\bs{1}^+)_R\oplus (\bs{3},\bs{1}^0)_R$  where the subscript denotes chirality and the superscript denotes electric charge. Subsequently, the $SU_{S}(3)$ environment induces $(\bs{3},\bs{2})_L\oplus (\bs{3},\bs{1}^+)_R\oplus (\bs{3},\bs{1}^0)_R\rightarrow (\bs{1},\bs{2})_L\oplus (\bs{1},\bs{1}^+)_R\oplus (\bs{1},\bs{1}^0)_R$. There are analogous transitions for the other representations. From \S \ref{trans-representation}, they are associated with the scattering operators $\bs{S}_0,\bs{S}_1,\bs{S}_2,\bs{S}_3$.

Recall that $\bs{S}_\alpha$ encodes $SU_S(3)$ coupling to its $U_{EW}(2)$ environment and vice versa for $\bs{S}_\sigma$, and they are responsible for the transitions
\begin{eqnarray}\label{transition diagram}
&&\bs{S}_0\stackrel{\bs{S}_\sigma}{\longrightarrow}
\bs{S}_2\stackrel{\bs{S}_\alpha}{\longrightarrow}\bs{S}_3\notag\\ \notag\\
&&\bs{S}_0\stackrel{\bs{S}_\alpha}{\longrightarrow}
\bs{S}_1\stackrel{\bs{S}_\sigma}{\longrightarrow}\bs{S}_3\;.
\end{eqnarray}
Clearly $\bs{S}_1$ acts on the iquark representation while $\bs{S}_3$ acts on the lepton representation. It is tempting to think that $\bs{S}_2$ governs new particle types appearing somewhere above the EW scale and likewise for $\bs{S}_0$. On the other hand, perhaps the transition represented by the top line does not occur.

The idea is that each scattering operator governs a subspace of phase space where only its associated matter field representation is manifest. The only gauge boson that mediates between these subspaces is the photon, because evidently it survives the decoherence. Gauge invariance then rests on anomaly cancelation, and this condition fixes the relative contribution of the matter fields in each subspace to the total Lagrangian. The possible Lagrangian content is tabulated below.
\begin{table}[h]
\centering
\begin{tabular}{|l|l|l|l|l|}\hline
 & \hspace{.1 in}$\bs{S}_0$  & \hspace{.8 in} $\bs{S}_1$ & \hspace{.4 in} $\bs{S}_2$  & \hspace{.8 in} $\bs{S}_3$  \\ \hline\hline
 $\kappa^{+}$ & $(\bs{3},\bs{2})$ &$(\bs{3},\bs{2})_L,(\bs{3},\bs{1}^+)_R, (\bs{3},\bs{1}^0)_R$  &$(\bs{1},\bs{2}),(\bs{1},\bs{2})$ & $(\bs{1},\bs{2})_L,(\bs{1},\bs{1}^+)_R,(\bs{1},\bs{1}^0)_R$ \\
 $\kappa^{-}$ & $(\bs{3},{\mathbf{2^c}})$ & $(\bs{3},{\mathbf{2^c}})_L,(\bs{3},\bs{1}^-)_R, (\bs{3},\bs{1}^0)_R$ & $(\bs{1},{\mathbf{2^c}}),(\bs{1},{\mathbf{2^c}})$ & $(\bs{1},{\mathbf{2^c}})_L,(\bs{1},\bs{1}^-)_R,(\bs{1},\bs{1}^0)_R$ \\
  $\bar{\kappa}^{+}$ & $(\overline{\bs{3}},\overline{\bs{2}})$ & $(\overline{\bs{3}},\overline{\bs{2}})_L,(\overline{\bs{3}},\bs{1}^-)_R, (\overline{\bs{3}},\bs{1}^0)_R$ & $(\bs{1},\bs{2}),(\bs{1},\bs{2})$ & $(\bs{1},\bs{2})_L,(\bs{1},\bs{1}^-)_R,(\bs{1},\bs{1}^0)_R$ \\
  $\bar{\kappa}^{-}$ & $(\overline{\bs{3}},\overline{{\mathbf{2^c}}})$ & $(\overline{\bs{3}},\overline{{\mathbf{2^c}}})_L,(\overline{\bs{3}},\bs{1}^+)_R, (\overline{\bs{3}},\bs{1}^0)_R$ & $(\bs{1},\overline{{\mathbf{2^c}}}),(\bs{1},\overline{{\mathbf{2^c}}})$ & $(\bs{1},\overline{{\mathbf{2^c}}})_L,(\bs{1},\bs{1}^+)_R,(\bs{1},\bs{1}^0)_R$ \\ \hline
\end{tabular}
\caption{Fermion field content for effective Lagrangian densities.}
\end{table}
\newline
To emphasize; the scale factors $\kappa^{+},\kappa^{-},\bar{\kappa}^{+},\bar{\kappa}^{-}$ (possibly different in respective phase space regions) are partly determined by anomaly cancellation, and not all terms in the table necessarily contribute to the Lagrangian.

Let's write down matter field contributions for each phase space region. For $\bs{S}_0$, let $\Bold{H}^{++}$ represent an element in the $(\bs{3},\bs{2})$ representation and $\D^{++}_0$ its associated covariant derivative. The superscripts refer to the sign of the strong and electric  charges (in that order). Assuming there are $s$ fermion generations,
\begin{eqnarray}\label{S0}
\bs{S}_0:&&\;\sum_{s}
\kappa_0^{+}\overline{\Bold{H}_s^{++}}\D_0^{++}
 \Bold{H}^{++}_s
 +\bar{\kappa}_0^{-}\overline{\Bold{H}_s^{--}}\D_0^{--}\Bold{H}^{--}_s\notag\\
&&\hspace{.3 in}+\bar{\kappa}_0^{+}\overline{\Bold{H}_s^{-+}}\D_0^{-+}
 \Bold{H}^{-+}_s+\kappa_0^{-}\overline{\Bold{H}_s^{+-}}\D_0^{+-}
 \Bold{H}^{+-}_s\notag\\\notag\\
&&\hspace{.0 in}=\sum_{s}
\overline{\Bold{H}_s^{++}}\D_0^{++}
 \Bold{H}_s^{++}
 +\overline{\Bold{H}_s^{--}}\D_0^{--}\Bold{H}_s^{--}\notag\\
&&\hspace{.3 in}+\overline{\Bold{H}_s^{-+}}\D_0^{-+}
 \Bold{H}_s^{-+}+\overline{\Bold{H}_s^{+-}}\D_0^{+-}
 \Bold{H}_s^{+-}\;.
\end{eqnarray}
A real-valued Lagrangian density requires $\kappa_0^{+}=\bar{\kappa}_0^{-}$ and $\kappa_0^{-}=\bar{\kappa}_0^{+}$. The second equality in (\ref{S0}) follows because the $SU_S(3)$ complex conjugate representations are inequivalent so we can arrange for $\kappa_0^{\pm}=\bar{\kappa}_0^{\pm}=1$ by suitable normalization of the carrying vector spaces.

The iquark contribution is
\begin{eqnarray}
\bs{S}_1:&&\;\sum_{s}
 {\kappa_1^{+}(\overline{\Bold{H}_{{\mathrm{L}},s}^{++}}\D_0^{++}
 \Bold{H}^{++}_{{\mathrm{L}},s}+\overline{\Bold{h}_{{\mathrm{R}},s}^{++}}\D_1^{++}
 \Bold{h}^{++}_{{\mathrm{R}},s}+\overline{\Bold{\xi}_{\mathrm{R,s}}^{+0}}\D_1^{+0}\Bold{\xi}^{+0}_{\mathrm{R},s})}\nonumber\\
 &&\hspace{.35in}
 +\kappa_1^{-}(\overline{\Bold{H}_{{\mathrm{L}},s}^{+-}}\D_0^{+-}
 \Bold{H}^{+-}_{{\mathrm{L}},s}+\overline{\Bold{h}_{{\mathrm{R}},s}^{+-}}\D_1^{+-}
 \Bold{h}^{+-}_{{\mathrm{R}},s}+\overline{\Bold{\chi}_{\mathrm{R,s}}^{+0}}\D^{+0}\Bold{\chi}^{+0}_{\mathrm{R},s})\notag\\\notag\\
&&\hspace{.35in}
+{\bar{\kappa}_1^{+}(\overline{\Bold{H}_{{\mathrm{L}},s}^{-+}}\D_0^{-+}
 \Bold{H}^{-+}_{{\mathrm{L}},s}+\overline{\Bold{h}_{{\mathrm{R}},s}^{-+}}\D_1^{-+}
 \Bold{h}^{-+}_{{\mathrm{R}},s}+\overline{\Bold{\xi}_{\mathrm{R,s}}^{-0}}\D_1^{-0}\Bold{\xi}^{-0}_{\mathrm{R},s})}\notag\\\notag\\
&&\hspace{.35in}
+{\bar{\kappa}_1^{-}(\overline{\Bold{H}_{{\mathrm{L}},s}^{--}}\D_0^{--}
 \Bold{H}^{--}_{{\mathrm{L}},s}+\overline{\Bold{h}_{{\mathrm{R}},s}^{--}}\D_1^{--}
 \Bold{h}^{--}_{{\mathrm{R}},s}+\overline{\Bold{\chi}_{\mathrm{R,s}}^{-0}}\D_1^{-0}\Bold{\chi}^{-0}_{\mathrm{R},s})} \end{eqnarray}
where $\D_1^{\pm\pm}=\mathrm{tr}_{U_{EW}(2)}\D_0^{\pm\pm}$ and $\D_1^{\pm0}$ is the restriction of $\D_0^{\pm\pm}$ to the trivial representation of $U_{EW}(2)$. Re-scaling the representation bases leaves the undetermined ratios $\kappa_1^{+}/\kappa_1^{-}$ and $\bar{\kappa}_1^{+}/\bar{\kappa}_1^{-}$ which are fixed by $\kappa_1^{+}+\kappa_1^{-}=1$, $\bar{\kappa}_1^{+}+\bar{\kappa}_1^{-}=1$, and anomaly cancellation as we have seen in \S\ref{SM comparison}.

Next, again a real-valued Lagrangian density implies $\kappa_2^{\pm}=\bar{\kappa}_2^{\mp}$ and inequivalent complex conjugate representations imply $\kappa_2^{\pm}=\bar{\kappa}_2^{\pm}$ so
\begin{eqnarray}\label{S2}
\bs{S}_2:&&\;\sum_{s}
\left(\overline{\Bold{h}_s^{++}}\,{{\D}_2^{++}}
 \Bold{h}^{++}_s+\overline{\Bold{h}_s^{0+}}\,{{\D}_2^{0+}}
 \Bold{h}^{0+}_s +\overline{\Bold{h}_s^{-+}}\,{{\D}_2^{-+}}
 \Bold{h}^{-+}_s+\overline{\Bold{h}_s^{0+}}\,{{\D}_2^{0+}}
 \Bold{h}^{0+}_s\right)\notag\\\notag\\
&&\hspace{.35 in} +\left(\overline{\Bold{h}_s^{--}}\,{{\D}_2^{--}}
 \Bold{h}^{--}_s+\overline{\Bold{h}_s^{0-}}\,{{\D_2^{0-}}}
 \Bold{h}^{0-}_s+\overline{\Bold{h}_s^{+-}}\,{{\D}_2^{+-}}
 \Bold{h}^{+-}_s+\overline{\Bold{h}_s^{0-}}\,{{\D}_2^{0-}}
 \Bold{h}^{0-}_s\right)
 \end{eqnarray}
where ${{\D}_2^{\pm\pm}}$ is $\mathrm{tr}_{SU_S(3)}\D_0^{\pm\pm}$, and ${\D}_2^{0\pm}$ (which contains no $SU_S(3)$ generator) is the covariant derivative in the trivial $SU_S(3)$ representation. Again we absorbed $\kappa_2^\pm$ into the $SU_S(3)$ complex conjugate  normalizations.

Finally, the lepton contribution is (there is no need to distinguish between $\kappa$ and $\bar{\kappa}$)
\begin{eqnarray}
\bs{S}_3:&&\;\sum_{s}
{\kappa}_3^{+}\left(\overline{\Bold{h}_{\mathrm{L},s}^{+}}\,{{\D}_2^{0+}}
 \Bold{h}^{+}_{\mathrm{L},s}+\overline{\Bold{h}_{\mathrm{R},s}^{+}}\,{{{\D}}_3^{+}}
 \Bold{h}^{+}_{\mathrm{R},s}+\overline{\Bold{\xi}_{\mathrm{R},s}^{0}}\,{{\D}_3^{0}}
 \Bold{\xi}^{0}_{\mathrm{R},s}\right)\notag\\
 &&\hspace{.3 in}+{\kappa}_3^{-}\left(\overline{\Bold{h}_{\mathrm{L},s}^{-}}\,{{\D}_2^{0-}}
 \Bold{h}^{-}_{\mathrm{L},s}+\overline{\Bold{h}_{\mathrm{R},s}^{-}}\,{{{\D}}_3^{-}}
 \Bold{h}^{-}_{\mathrm{R},s}+\overline{\Bold{\chi}_{\mathrm{R},s}^{0}}\,{{\D}_3^{0}}
 \Bold{\chi}^{0}_{\mathrm{R},s}\right)
 \end{eqnarray}
 where $\D_3^{-}=\mathrm{tr}_{U_{EW}(2)}\D_2^{0-}$ and $\D_3^0=\partial\!\!\!/$. Experiment dictates $\kappa_3^-=1$ and $\kappa_3^+=0$ so
 \begin{eqnarray}
\bs{S}_3:&&\;\sum_{s}
\overline{\Bold{h}_{\mathrm{L},s}^{-}}\,{{\D}_2^{0-}}
 \Bold{h}^{-}_{\mathrm{L},s}+\overline{\Bold{h}_{\mathrm{R},s}^{-}}\,{{{\D}}_3^{-}}
 \Bold{h}^{-}_{\mathrm{R},s}+\overline{\Bold{\chi}_{\mathrm{R},s}^{0}}\,{{\D}_3^{0}}
 \Bold{\chi}^{0}_{\mathrm{R},s}\;.
 \end{eqnarray}

Given this matter field content, we need to verify there are no anomalies. The $\bs{S}_0$ and $\bs{S}_2$ sectors exhibit chiral symmetry so they are safe. For the remaining $\bs{S}_1$ and $\bs{S}_3$ sectors, the case is already settled; no anomaly if $\kappa_1^{+}=\bar{\kappa}_1^{+}=2/3$ and $\kappa_1^{-}=\bar{\kappa}_1^{-}=1/3$.

\end{document}